\documentclass[11pt, a4paper]{article}

\usepackage[a4paper, left=3.0cm, right=3.0cm, top=2.5cm, bottom=2.5cm,]{geometry}

\usepackage[english]{babel}
\usepackage{amsmath}
\usepackage{amssymb}
\usepackage{mathabx}
\usepackage{graphics}
\usepackage{graphicx}
\usepackage{fancyhdr}
\usepackage{setspace}
\usepackage[modulo]{lineno}
\usepackage{lastpage}
\usepackage{lmodern}
\usepackage{comment}
\usepackage{tabularx}
\usepackage{booktabs}
\usepackage{multirow}
\usepackage{subcaption}
\usepackage{bm}
\usepackage{color}
\usepackage{units}
\usepackage{float}
\usepackage{xcolor}
\usepackage[table]{xcolor}
\usepackage{datetime}
\usepackage{array}
\usepackage{pdflscape}	
\usepackage{blindtext}
\usepackage{textcomp}
\usepackage[gen]{eurosym}
\usepackage{relsize}
\usepackage{rotating}
\usepackage{leftindex}

\usepackage{titlesec}
\titleformat{\section}{\large\bfseries}{\thesection}{1em}{}

\usepackage{subfiles}
\usepackage{booktabs}
\usepackage{dcolumn}
\usepackage{multirow, makecell, caption}

\usepackage[round]{natbib} 
\newcounter{savefootnote}
\newcounter{symfootnote}
\newcommand{\symfootnote}[1]{%
	\setcounter{savefootnote}{\value{footnote}}%
	\setcounter{footnote}{\value{symfootnote}}%
	\ifnum\value{footnote}>8\setcounter{footnote}{0}\fi%
	\let\oldthefootnote=\thefootnote%
	\renewcommand{\thefootnote}{\fnsymbol{footnote}}%
	\footnote{#1}%
	\let\thefootnote=\oldthefootnote%
	\setcounter{symfootnote}{\value{footnote}}%
	\setcounter{footnote}{\value{savefootnote}}%
}

\titlespacing\section{0pt}{3ex plus 1ex minus 0ex}{.5ex minus .5ex}
\titlespacing\subsection{0pt}{3ex plus 1ex minus 0ex}{.5ex plus .2ex}
\titlespacing\subsubsection{0pt}{2ex plus 1ex minus 0ex}{.5ex plus .2ex}

\author{\myauthor}
\usepackage[colorlinks,linkcolor=blue,citecolor=blue,urlcolor=blue]{hyperref} 
\addto\captionsenglish{%
}

\usepackage[normalem]{ulem}

\newenvironment{keywords}{\noindent\textbf{Keywords:}}{}

\newcommand{\git}{\href{https://git.rwth-aachen.de/avt-svt/public/representative-electricity-price-profiles}{\texttt{Link}}}

\newcommand{\mytitle}{A practical scenario generation method for electricity prices on day-ahead and intraday spot markets}

\begin{document}

\newpage
	
\begin{center}
	\begin{Large}
		\textbf{\mytitle}
	\end{Large}

    \vspace{2ex}
	Chrysanthi Papadimitriou$^{a,}$\symfootnote{Both authors contributed equally to this work},
    Jan C.~Schulze$^{a,*}$,
    Alexander Mitsos$^{c,a,b,}$\symfootnote{Correspondence: \texttt{amitsos@alum.mit.edu} }
\end{center}

\vspace{-2ex}
\begin{scriptsize}
\begin{center}
$^a$ Process Systems Engineering (AVT.SVT), RWTH Aachen University, 52074 Aachen, Germany \\
$^b$ Energy Systems Engineering (ICE-1), Forschungszentrum J\"ulich, 52425 J\"ulich, Germany \\
$^c$ JARA-ENERGY, 52056 Aachen, Germany \\
\end{center}
\end{scriptsize}


\vspace{-3ex}
\begin{center}\textbf{Abstract}\end{center}
\vspace{-5ex}
The increasing interest in demand-side management (DSM) as part of the energy cost optimization calls for effective methods to determine representative electricity prices for energy optimization and scheduling investigations.
We propose a practical method to construct price profiles of day-ahead (DA) and intraday (ID) electricity spot markets.
We construct single-day and single-week price profiles based on historical market time series to provide ready-to-use price data sets.
Our method accounts for dominant mechanisms in price variation to preserve critical statistical features (e.g., mean and standard deviation) and transient patterns in the constructed profiles. 
Unlike common scenario generation approaches, the method is deterministic, with few degrees of freedom and minimal application effort.
Our method ensures consistency between ID and DA price profiles when both are considered and introduces profile scaling to enable multiple scenario generation.
Finally, we compare the constructed profiles to clustering techniques in a DSM case study, noting similar cost results.

\begin{keywords}
    Pricing,
    Energy markets,
    Double peak profile,
    Demand response,
    Flexible operation
\end{keywords}

{
	\footnotesize
	Published in \textit{Computers and Chemical Engineering}:
	\href{https://doi.org/10.1016/j.compchemeng.2025.109118}{DOI:10.1016/j.compchemeng.2025.109118}
	\textcopyright{} 2025. This manuscript version is made available under the CC-BY 4.0 license (\url{https://creativecommons.org/licenses/by-nc-nd/4.0/}).\\		
}

\vspace{-3ex}
\section{Introduction}\label{sec:Introduction}
\vspace{-3ex}
Optimization of energy costs has long been acknowledged as a topic of interest for the economic design and operation of chemical processes and energy systems \citep{Friedler.2010}. 
Demand-side management (DSM) and demand response (DR) are energy-flexible operation paradigms that have have raised interest of industrial consumers to actively participate in the energy spot markets by adjustment of electricity demand \citep{Pinson.2014, Mitsos.2018}. 
With electricity prices on European DA spot markets having reached peak values as high as \unit[4\,000]{\euro/MWh} \citep{NordPool},
the importance of industrial DSM is expected to grow further in the near future.
Thus, economic studies on planning and scheduling of industrial processes necessitate a systematic way to select up-do-date electricity price scenarios. 
On the other hand, the process of selecting a price scenario should not distract from the main purpose of a scientific study on DSM.
Thus, we herein propose a user-friendly intuitive approach to determine standard price profiles.
Notice that we use the terms ``price profiles'' and ``prices time series'' interchangeably.

Publications on DSM of industrial processes, 
including our own studies, such as \cite{Caspari.2019,Mucci.2023,Elwajeh.2023}, 
heavily rely on the chosen electricity price profiles that must represent current trends and allow for generalizable conclusions. 
Selecting a single historical profile or averaging historical data favors simplicity, and is essential in cases of computationally expensive calculations (e.g., \cite{Papadimitriou.2023}), or a case study number that scales significantly with the number of price scenarios considered (e.g., \cite{Kelley.2020c}).
Clearly, averaged price trends do not fully capture all features of historical profiles, as price averaging yields a single and exclusive price scenario.
Additionally, data averaging does not account for the temporal evolution of prices to consider future price trends.
Nonetheless, simple and sufficiently representative price profiles are usually adequate for evaluation of the theoretical potential of DR applications.
Conversely, sophisticated scenario generation methods provide multiple price scenarios.
These scenarios are rather valuable for active market participation, though less suitable for generalizable results on the considered DR method.
Yet, these methods involve a more complex implementation, while forecasts of the long-term price trends are still subject to high uncertainty.
Moreover, an extensive evaluation of all possible price scenarios, product demand situations, and other possible operating disturbances quickly results in a curse of dimensionality and is thus often beyond the scope of research studies on DR.

We propose a price scenario generation method, by considering two key price components of widely considered DR, namely the auction day-ahead (DA) and the real-time intraday (ID) electricity spot market.
In order to decide on the time horizon of the generated price profiles, we note that in most related studies single-day or single-week price profiles are utilized, while less frequently longer horizons of historical data (e.g., up to one year) may be applied  \citep{Meese.2016}.
The daily and weekly frequencies match the dominant patterns of the DA and ID price fluctuation identified in our previous studies (\cite{Schafer.2020,Germscheid.2022}).
We can thus extend the previously introduced averaging techniques to construct single-day and single-week DA and ID electricity price profiles.

Herein, we aim to establish a simple method to construct meaningful and representative DA and ID ready-to-use price profiles to be used within future economic studies, e.g., 
cost optimal planning and scheduling, 
power-to-X, 
and energy-flexible design and operation.
Within this method, we desire an option to allow consistency of the DA and ID prices, which is useful for DR across multiple electricity markets, e.g., in bidding strategies \citep{Silva.2022}, or scheduling and control \citep{Caspari.2020}.
This ensures that the absolute DA and ID price values of a selected price scenario do not na\"ively favor a certain DR strategy.
An example of the effect of a systematic shift of DA and ID prices is given in our recent work \cite{Schulze.2024}, where flexible operation in the more volatile ID market appears less profitable than in the DA market, because of higher ID price values.

This property is not provided by existing methods, e.g., \cite{Maciejowska.2019,Teichgraeber.2019}.
The developed price profile will exhibit the following essential properties:
\begin{enumerate}
\vspace{-2ex}
\setlength\itemsep{0ex}
    \item Method simple to understand, deterministic, and few parameters, \label{req:simple}
    \item Evolution of electricity prices closely represents current price trends,\label{req:reality}    
    \item Consistency of DA and ID price profiles, \label{req:comarable}
    \item Magnitude of price fluctuations is adjustable (scalable), while 1.~-~3.~still hold. \label{req:scalabe}
\end{enumerate}

To achieve the desired properties, we base our method on previous works on constructing average price profiles \citep{Schafer.2020, Germscheid.2022} and introduce adaptations tailored to the applications investigated.
Because the focus of DR studies is on the economic potential of electricity price fluctuations rather than steady-state optimization, the price variance over the investigation period is assumed to hold greatest importance for optimal economic decision-making \citep{Mathaba.2014}.
Thus, our method allows for adjusting the magnitude of fluctuations in the price profile.
Additionally, we ensure consistency of the DA and ID scenarios based on the near-zero mean distribution of the ID-DA deviation by equalizing the integrated DA and ID prices over the designated period.
To underpin this approach, we provide  statistical analyses of hourly DA and quarter-hourly real-time ID prices in historical data from European spot markets. 
Our approach can extend to other markets and price granularities (e.g., half-hourly) where similar observations hold.

In the remainder of this article, we first present state-of-the-art approaches on electricity price scenario generation for DSM (Section \ref{sec:SOA}). 
Subsequently, we introduce a method for constructing a practical 24-hour DA price profile from historical data of the last available 365 days (Section \ref{sec:DAconstruction}). 
Next, we utilize the same historical data to extract the ID price profile, based on ID - DA price deviations (Section \ref{sec:IDconstruction}). 
Then, we introduce a few modifications and apply the same approach to generate single-week (Section \ref{sec:priceprofileS1723_week}) practical profiles. 
We further create ready-to-use\footnote{The open-access data sets are provided via Git (\git).}
DA and ID price profiles for future studies on DSM on the different time-scales presented and compare to historical data (Section \ref{sec:Results}). 
Last, we present a DR case study to contrast our method and state-of-the-art price generation techniques in terms of cost results (Section \ref{sec:compSOTA}).

\vspace{-1ex}
\section{State of the art}\label{sec:SOA}
\vspace{-3ex}
Next, we review state-of-the-art techniques for electricity price scenario selection or generation.
An overview of the presented methods is shown in Table~\ref{tab:SOTA}, together with exemplary works of electricity price consideration.
While not exhaustive, the presented classification covers the most common price generation methods and aligns with the objectives of our framework.

The most straightforward approach is to use recent historical profiles as scenarios.
Several studies consider a single historical profile, e.g., 24 hours, that exhibits a typical behavior in terms of the price range and transient patterns \citep{DalleAve.2019}. 
The authors then expect the findings to generalize well.
Other works consider several different historical profiles commonly representing different extreme scenarios in their DSM application to highlight a range of the expected results \citep{Caspari.2020,Bree.2019}.
Especially when studies on production design and planning are performed, a higher number of historical price profiles can be selected to capture the effect of different fluctuation patterns \citep{Fursch.2013}.
The above scenario generation strategies are based on the use of raw historical data, yielding realistic results of the DSM application while keeping the number of scenarios low. 
However, they usually proceed ad-hoc, resulting in challenges for the representativity and generalizability of case study results.

A more sophisticated approach to determine characteristic periods in time-series data is time series aggregation (TSA) \citep{Hoffmann.2020}.
Within this category, the clustering of historical price data is a commonly used technique. 
Historical price clustering provides a set of price profiles that represent the entity of the historical data \citep{Poncelet.2017}. 
All clustering techniques involve three core steps \citep{Teichgraeber.2019}.
First, clustering criteria are defined to span the clustering space.
For high-dimensional data vectors, e.g., price profiles,
common low-dimensional clustering criteria include the mean and standard deviation of the data point \citep{Wang.2006}.
The vector of clustering criteria then represents the clustering candidates associated to original data vectors.
Second, the distance between candidate cluster members and candidate cluster center is minimized to determine clusters and associated center points.
Third, the cluster representative (characteristic profile) is derived from the cluster members.

The most common clustering strategies are $k$-means clustering, $k$-medoids clustering, and hierarchical clustering.
A $k$-means clustering starts by randomly selecting $k$ cluster centers.
Each cluster center is the mean, termed centroid, of the cluster members assigned to the cluster.
Further, the cluster representative price profile is the mean of the profiles corresponding to the cluster \citep{Jain.2010}.
In $k$-medoids clustering, the most central cluster member, called medoid, is the center and derives the representative profile \citep{Kaufman.2009}.
Hierarchical clustering begins with single-member clusters that are iteratively merged using centroids as cluster centers.
The cluster representatives are either defined as medoids or centroids \citep{Nahmmacher.2016, Teichgraeber.2019}.
\cite{Kotzur.2018} compare different clustering techniques for price scenario generation applied to optimal process design.
Furthermore, \cite{Wang.2006} explore different clustering criteria to enhance clustering performance.

Although clustering reduces the number of scenarios compared to the original time series, multiple clusters within a DR study require multiple scheduling problems to be solved.
Clustering techniques provide a more systematic way for historical profile selection compared to applying user-specified criteria with high degree of freedom.
However, numerous options exist for configuring clustering methods.
One such option is data scaling, which is essential for time-series clustering and can significantly impact results \citep{Rager.2015}, but is not required for averaging and moment-matching techniques.
Yet, the clustering results are only optimal with respect to the chosen criteria. In other words, the price profiles associated with cluster centers are the best representatives with respect to the chosen criteria, but they do not necessarily yield good results when applied to a DR case study.
This can occur when employing too few criteria and risking loss of critical information.
Conversely, using too many criteria can dilute specificity and produce profiles far from the average expected performance.

In an effort to enhance clustering performance, hybrid techniques that combine time-series feature extraction, such as principal component analysis (PCA), together with (machine learning) clustering techniques have been introduced for similar applications (e.g., load profiles) to capture underlying patterns and improve clustering \citep{Almaimouni.2018}. 
Besides PCA, factor analysis methods are used to uncover latent factors and explain data variability.
Examples include latent Dirichlet allocation \citep{Chen.2022} and autoencoder networks, e.g., variational autoencoders \citep{Wang.2022b}.
Similarly, generative AI approaches like generative adversarial networks \citep{Yilmaz.2024} or jump-diffusion models \citep{Meyer-Brandis.2008} can produce realistic data by learning the underlying data distribution.
These methods can capture complex, nonlinear relationships and generate diverse, realistic profiles.
However, they require increased computational complexity and careful tuning compared to moment-matching techniques.

An alternative way to produce practical price scenarios using historical data is to construct average price profiles over a historical period that respects fundamental characteristics of the spot market price profiles.
\cite{Knittel.2005} identify electricity price trends on different time scales highlighting recurring patterns and characteristic values of the price time series. 
\cite{Rahimiyan.2016} perform a statistical analysis on historical DA and ID price data to quantify the price correlations.
\cite{Roder.2024} construct an average DA profile through permutation of the price elements of historical profile data together with averaging applicable to flexible operation for price peak avoidance.
\cite{Schafer.2020} and \cite{Germscheid.2022} utilize Fourier transforms to analyze principal frequencies of historical DA prices and ID-DA price deviations, respectively, 
identifying dominant modes at periods of \unit[0.5]{day} and \unit[1]{day}.
By averaging one year of historical DA and ID price data, significant time-series frequencies, such as daily or weekly, are taken into account to construct a single profile. 
These periods correspond to significant price patterns and price fluctuations that can be exploited by most processes. 
The averaging method is straightforward to implement compared to clustering techniques and provides a single scenario.
However, the approach may over-smooth the data and potentially obscure certain less prominent trends.

Finally, we mention probabilistic forecasting techniques, which exploit correlations of predicted parameters (e.g., weather forecasts) and are widely used to explain price behavior \citep{Nowotarski.2018}. 
Following \cite{Monteiro.2018} and \cite{Weron.2014}, these methods can be categorized into prediction interval, density, and threshold forecasting.
Univariate and multivariate probabilistic models are prevalent electricity price forecasting methods \citep{Cuaresma.2004,Rahimiyan.2016}, but machine learning approaches like normalizing flows are also gaining attention \citep{Cramer.2023}.
Electricity prices forecasts enable DR computations in future applications, offering advantages over historical data, especially for long horizons and volatile markets \citep{Nowotarski.2018}.
Unlike data-averaging techniques, price forecasting methods are not restricted to representing data of the present and recent past.
Instead, forecasting can incorporate trends of the future, although commonly restricted to the use of historical data and small forecasting windows \citep{Alonso.2016,Shah.2021}.
Notably, however, both historical data based as well as complex scenario forecasting methods have been rather inaccurate at predicting the long-term evolution of electricity prices \citep{Zareipour.2010, Weron.2014, Gabrielli.2022}, since this evolution is subject to policy changes and political events \citep{Yang.2017}.
At the same time, although practical for real-life application of DSM strategies, such as optimal bidding in a wholesale energy market \citep{Zhang.2020}, price forecasts refer to a very specific time period in the future.
They may therefore not provide a basis for generalizable conclusions when assessing the potential of an introduced DR case study.

In conclusion, we find all methods to be either fairly involved or subject to individual user decisions or randomness, i.e., non-deterministic.
Hence, we aim to provide a simple yet systematic and deterministic method for constructing single-day and single-week electricity price scenarios based on historical time series data.
Such a method shall involve few tuneable parameters and request little implementation effort.
In the next section, we propose such a method based on averaging.

\newcommand\mc[1]{\multicolumn{1}{l}{#1}}
\newcommand\mr[2]{\multicolumn{1}{l}{\multirow{#1}{*}{\makecell{#2}}}}

\begin{table}[h!bt]
	\centering
	\caption{Overview of the presented state-of-the-art techniques for electricity price scenario consideration. We distinguish between ``historical'' (real data) and ``artificial'' (synthetically generated) profiles, ``multiple'' or ``single'' number of generated profiles, and ``past'' or ``future'' represented time windows or price realization.}
	\label{tab:SOTA}
\begin{small}
{
\begin{tabular}{l llllll}
    \toprule
    \multirow{3}*{\textbf{Technique}} & \mr{2}{\textbf{Key}} & \textbf{Derived} & \textbf{Derived} & 
    \\ 
    & \mr{2}{\textbf{features}} & \textbf{price} & \textbf{profile} & \textbf{References}\\

     &  & \textbf{datasets} & \textbf{window} & \\
    \midrule
    \multirow{4}*{Recent prices} & price data & \mr{2}{historical,} & \multirow{4}*{past} & \cite{DalleAve.2019},\\
    & qualitative/ & \mr{2}{single/} & & \cite{Caspari.2020},\\
    & quantitative & \mr{2}{multiple} & & \cite{Bree.2019},\\
    & selection& & & \cite{Fursch.2013}\\
    \midrule
    \multirow{3}*{Clustering}& $k$-means, & historical/ & \multirow{3}*{past} & \cite{Kotzur.2018},\\
     & $k$-medoids, & artificial, & & \cite{Poncelet.2017},\\
     & hierarchical & multiple & & \cite{Teichgraeber.2019} \\
     \midrule
    \mr{2}{Hybrid} & feature & historical/  & \multirow{3}*{past} & \cite{Almaimouni.2018}\\
    \mr{2}{clustering} & extraction & artificial, & &\cite{Yilmaz.2024}\\ 
     & \& clustering & multiple & &\cite{Meyer-Brandis.2008}\\
    \midrule
    \multirow{3}*{Averaging} & of one-year & \mr{2}{artificial,}  & \multirow{3}*{past} & \cite{Roder.2024}\\
     &  historical & \mr{2}{single} & & \cite{Germscheid.2022}\\
     &  data &  & & \cite{Schafer.2020}\\
    \midrule
     & prediction &  & \multirow{6}*{future} & \cite{Nowotarski.2018}\\
      &  interval, &  & & \cite{Monteiro.2018} \\
     Probabilistic & density, & artificial, & & \cite{Weron.2014,Zhang.2020}\\
     forecasting & threshold, & single & & \cite{Cramer.2023}\\
      & univariate, &  & & \cite{Cuaresma.2004}\\
      & multivariate &  & & \cite{,Rahimiyan.2016}\\
    \bottomrule
\end{tabular}
}
\end{small}
\end{table}

\vspace{-1ex}
\section{Single-day price profile}\label{sec:OneDayPriceConstruction}
\vspace{-1ex}
In this section, we present the methodology for constructing representative single-day DA (Section \ref{sec:DAconstruction}) and ID (Section \ref{sec:IDconstruction}) price profiles using a full-year time-series data set.

\subsection{DA price profile}
\label{sec:DAconstruction} 
\vspace{-1ex}
We begin with the construction of the DA profile of one representative day. 
To this end, we select a price data set of 365 consecutive days to exclude seasonal effects \citep{Knittel.2005}.
The 365 days horizon may be chosen arbitrarily and if desired extended to multiple years.
We remark that incidents and trends contained in the price data set, e.g., caused by political events and developments, will affect the price scenario created. 
Here, we use up-to-date price values of the full year 2023. 
Similar to our prior works \citep{Schafer.2020,Germscheid.2022}, we first average the historical DA prices for every day hour, $k=1,2,...,24$, individually\footnote{
Notice that for $k\geq2$, Eq.~\eqref{eq:averageDA} may be equivalently rewritted as
$\overline{DA}_{k} = \overline{DA}_{k-1} + \overline{\Delta DA}_{k}$,
where
$\overline{\Delta DA}_{k} = 1/N \sum_{i=1}^{N}{ (DA_{i,k} - DA_{i,k-1})}$.
Hence, this profile captures the moving average price development over a day.
}:
\begin{equation}
    \label{eq:averageDA}
	\overline{DA}_{k} = \frac{1}{N} \sum_{i=1}^{N}{DA_{i,k}}
    \,,
\end{equation}
where $N = 365$ is the number of days of selected historical data,
$DA_{i,k}$ is the DA price at hour $k$ of day $i=1,2,...,N$,
and $\overline{DA}_k$ is the average DA price of hour $k$.
If desired, days with significant outliers, weekends or holidays may be excluded.
However, here we consider the full year.

\begin{figure}[ht]
	\centering    
    \includegraphics[width=0.7\linewidth]{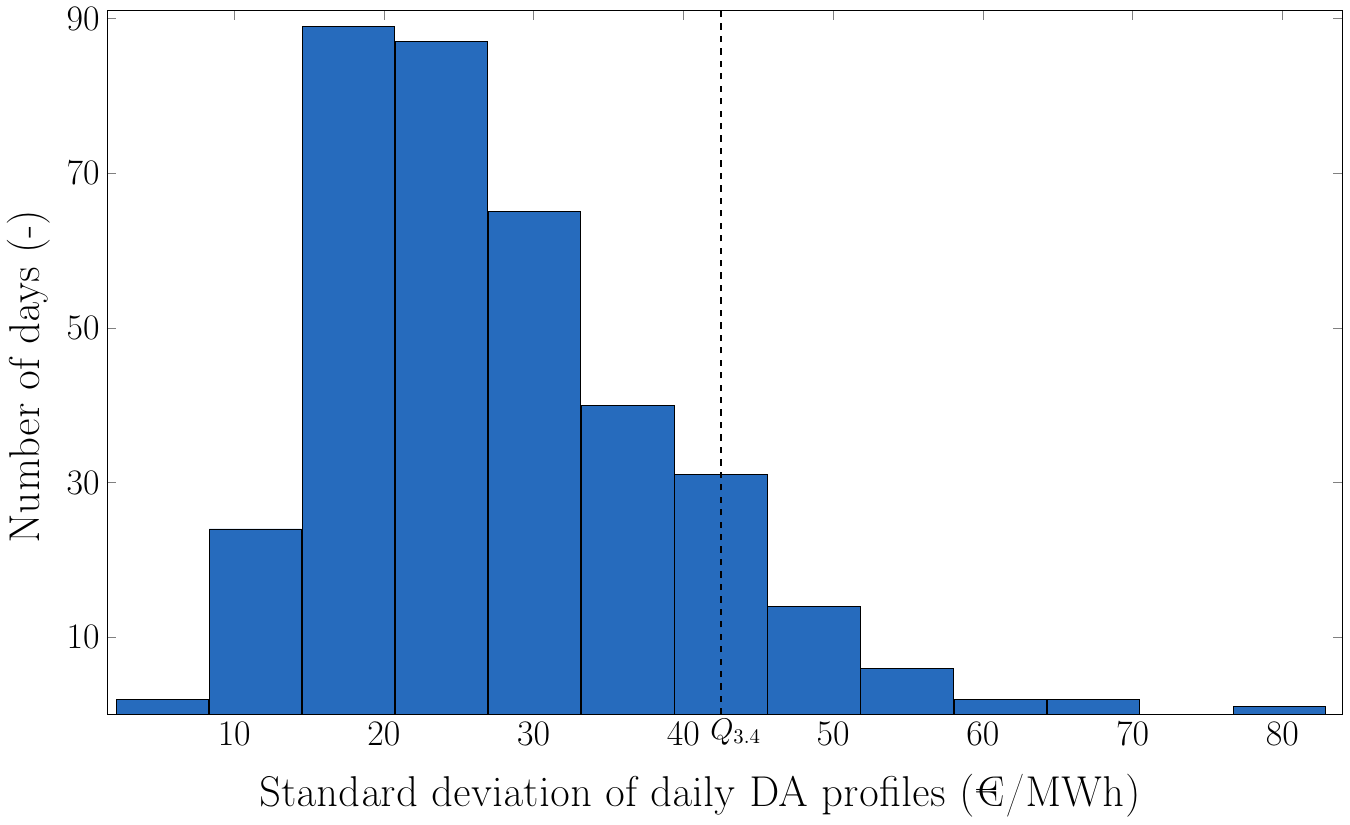}
	\caption{Distribution histogram of daily DA price standard deviation. Based on historical data of year 2023.}
	\label{fig:HistDAstd}
\end{figure}

Next, we calculate the overall mean value $\overline{DA}$:
\begin{linenomath}\begin{equation*}
\label{eq:meanofDA}
    \overline{DA} = \frac{1}{24} \sum_{k=1}^{24} \overline{DA}_{k} \,,
\end{equation*}\end{linenomath}
which is also the mean over all historical data.
\begin{figure}[tb]
	\centering
    \includegraphics[width=0.7\linewidth]{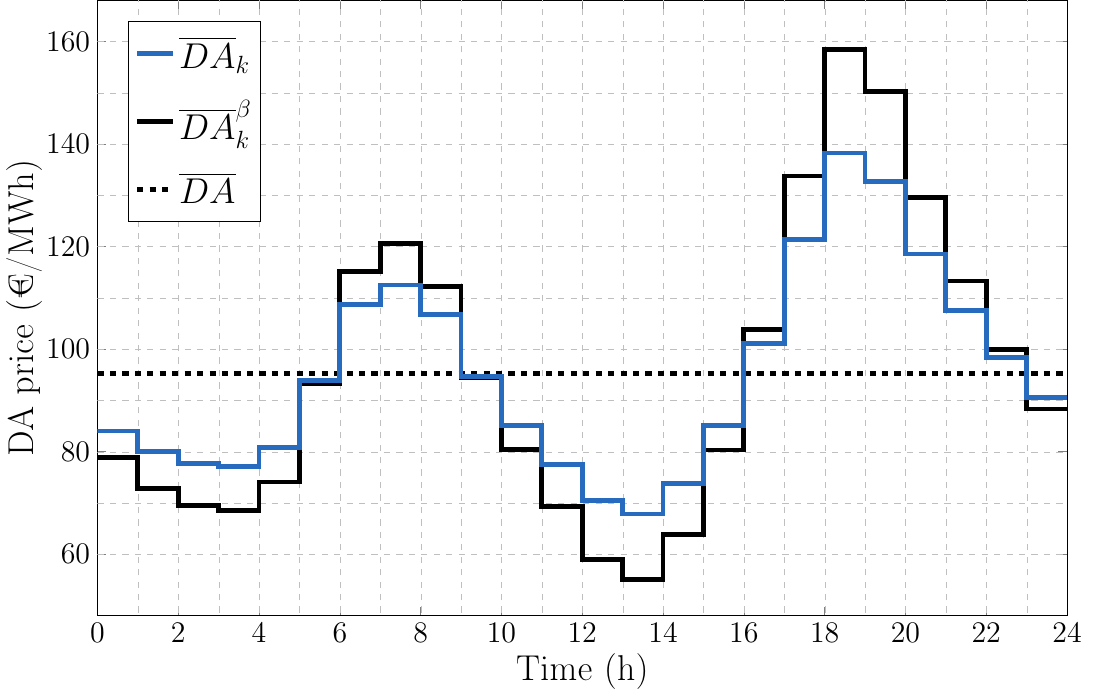}
	\caption{Average (\textit{Unscaled}) and scaled average ($\beta=1.47$) (\textit{Nominal}) 24-hour DA price profile generated from historical 2023 data \citep{EPEX}.}
	\label{fig:DA}
\end{figure}
To facilitate the adjustment of the price fluctuation around the mean value (see Requirement \ref{req:scalabe}), 
i.e., variance scaling,
we introduce a scaling factor $\beta > 0$.
This aspect extends our prior works.
We scale all $\overline{DA}_{k}$ around $\overline{DA}$ by $\beta$:
\begin{equation}
\label{eq:DApricescaling}
	\overline{DA}_{k}^{\beta} = \overline{DA} + \beta \cdot (\overline{DA}_{k} - \overline{DA})
    \,.
\end{equation}
Clearly, setting $\beta=1$ corresponds to no scaling and practically represents the centroid profile derived by any clustering technique which applies a single cluster and derivation of the representative profiles with centroids.
The resulting scenario exhibits low deviating behavior according to the historical data (Fig.~\ref{fig:HistDAstd}).
It could thus serve as a price scenario of moderate fluctuations.
Such an averaging-out of the data variation is also observed in centroid clustering methods, like $k$-means clustering \citep{Kotzur.2018,Schutz.2018}.
To minimize scenario complexity, we can thus use the $\beta=1$-profile as an \textit{Unscaled} scenario.
Importantly, the mean value $\overline{DA}$ remains identical irrespective of scaling.

Therefore, as the daily standard deviation of the average profile does not represent the daily standard deviation of all data due to smoothing of the data \citep{Schafer.2020}, we use $\beta$ to correct the averaged results. 
This allows to preserve a key time-series property critical for DR applications, namely the magnitude of price deviation. 
Thus, in order to obtain a \textit{Nominal} ready-to-use profile, 
we choose $\beta$ such that:
\begin{linenomath}\begin{equation*}
\label{eq:stdDAaverage}
  \sigma(\overline{DA}_{1}^{\beta}, ..., \overline{DA}_{24}^{\beta}) 
  = \frac{1}{N} \sum_{i=1}^{N} \sigma(DA_{i,1}, ..., DA_{i,24}) \,,
\end{equation*}\end{linenomath}
i.e., the standard deviation $\sigma$ of the resulting scaled price profile
matches the average standard deviation of the daily DA price profiles.
We insert the definition of the standard deviation:
\begin{linenomath}\begin{equation*}
    \sqrt{
    \frac{1}{24}
    \sum_{k=1}^{24}(\overline{DA}_{k}^\beta - \overline{DA})^2
    } =
    \frac{1}{N}
    \sum_{i=1}^{N}
    \sqrt{
    \frac{1}{24}
    \sum_{k=1}^{24}
    \big( DA_{i,k} - 
    \frac{1}{24}\sum_{j=1}^{24}DA_{i,j}
    \big)^2
    }
\end{equation*}\end{linenomath}
and obtain an explicit solution after mathematical reformulation:
\begin{linenomath}\begin{equation*}
    \label{eq:scalingbeta}
	\beta =
    \frac{1}{N}
    \sum_{i=1}^{N}
    \sqrt{
    \frac{
    \sum_{k=1}^{24}
    \big( DA_{i,k} - 
    \frac{1}{24}\sum_{j=1}^{24}DA_{i,j}
    \big)^2
    }{
    \sum_{k=1}^{24}(\overline{DA}_{k} - 
    \overline{DA} 
    )^2
    }}
    \,.
\end{equation*}\end{linenomath}
For example, for the full year 2023 and EPEX spot market, we receive the value $\beta = 1.47$.
The resulting DA profile is illustrated by Fig.~\ref{fig:DA}.
Here, the characteristic double peak profile is visible with minimal values occurring during the 4th and 14th hour
as well as maximal values during the 8th and the 19th.
The absolute price range after scaling ($\beta=1.47$) is $\unit[103.46]{\euro/MWh}$
with a mean value of $\overline{DA}=\unit[95.18]{\euro/MWh}$.

If a more distinct price variation is desired, we can specify a larger value of $\beta$. 
In order to systematically choose a larger $\beta$ factor that corresponds to some extreme behavior of the historical data (Requirement \ref{req:reality}), we represent the density distribution of the standard deviations $\sigma(DA_{i,1}, ..., DA_{i,24})$ across all days $i = 1, 2, ..., N$ through a histogram, Fig.~\ref{fig:HistDAstd}. 
Subsequently, we identify a quantile indicative of high fluctuations within the historical data. 
We exemplary show the quantile 3.4 ($Q_{3.4}$), corresponding to the 15\% highest values of the daily standard deviations, Fig.~\ref{fig:HistDAstd}. 
The value of the quantile is the corresponding standard deviation, namely:
\begin{linenomath}\begin{equation*}\label{eq:quantile}
  \sigma(\overline{DA}_{1}^{\beta}, ..., \overline{DA}_{24}^{\beta}) 
  = Q_{3.4}\big(\{\sigma(DA_{i,1}, ..., DA_{i,24})\}_{i=1}^N\big) \,,
\end{equation*}\end{linenomath}
satisfied for:
\begin{linenomath}\begin{equation*}
    \label{eq:scalingbetaextreme}
	\beta =
    \frac{Q_{3.4}\big(\{\sigma(DA_{i,1}, ..., DA_{i,24})\}_{i=1}^N\big)}{
    \sqrt{
    \frac{1}{24}
    \sum_{k=1}^{24}(\overline{DA}_{k} - 
    \overline{DA})^2
    }}
    \,,
\end{equation*}\end{linenomath}
giving $\beta=1.97$ for the full year 2023. 
Other options could include the choice of a preset high value, e.g., as an expression of the scaling parameter $\beta$ that corresponds to the \textit{Nominal} profile.
Conversely, for a less aggressive profile, we may set $\beta<1$ to match, e.g., $Q_{0.8}$ (of the 20\% least deviating historical data). 
The use of such variations on the averaged profile variance enables various different scenarios to be used in a DR case study.
Examples are optimal design and scheduling, where a sensitivity analysis over the prices allows to determine limits in equipment oversizing (e.g., \cite{Roder.2024}).

\vspace{-1ex}
\subsection{ID price profile}
\label{sec:IDconstruction}
\vspace{-1ex}
We proceed with generating an average ID profile by first exploring the relationship between DA and ID prices, followed by the construction of an ID profile based on a DA reference.
\cite{Rahimiyan.2016} assessed the correlation between DA and ID prices, pinpointing uncorrelated DA price and market deviation (also referred to as DA-ID price deviation or residual price). As noticed by \cite{Germscheid.2022}, the ID price oscillates around the DA price in a well-defined, season-independent harmonic pattern with dominant frequencies at $\unit[0.5]{h^{-1}}$ and close to $\unit[1]{h^{-1}}$.
We perform a similar analysis for the DA prices and the ID-DA deviation of 2023 (frequency spectra provided in the SI) that confirms the observations of \cite{Germscheid.2022}.
Consequently, ID profiles may be constructed by first subtracting ID and DA data and then using this market deviation to construct a harmonic deviation profile.
Here, we build on these findings and introduce some further refinements.
We consider the $\text{ID}_\text{3}$ price index, which is the volume-weighted average price of all trades that took place within the last three hours before delivery \citep{EPEX}. 

\begin{figure}[t]
	\centering
    \includegraphics[width=0.8\linewidth]{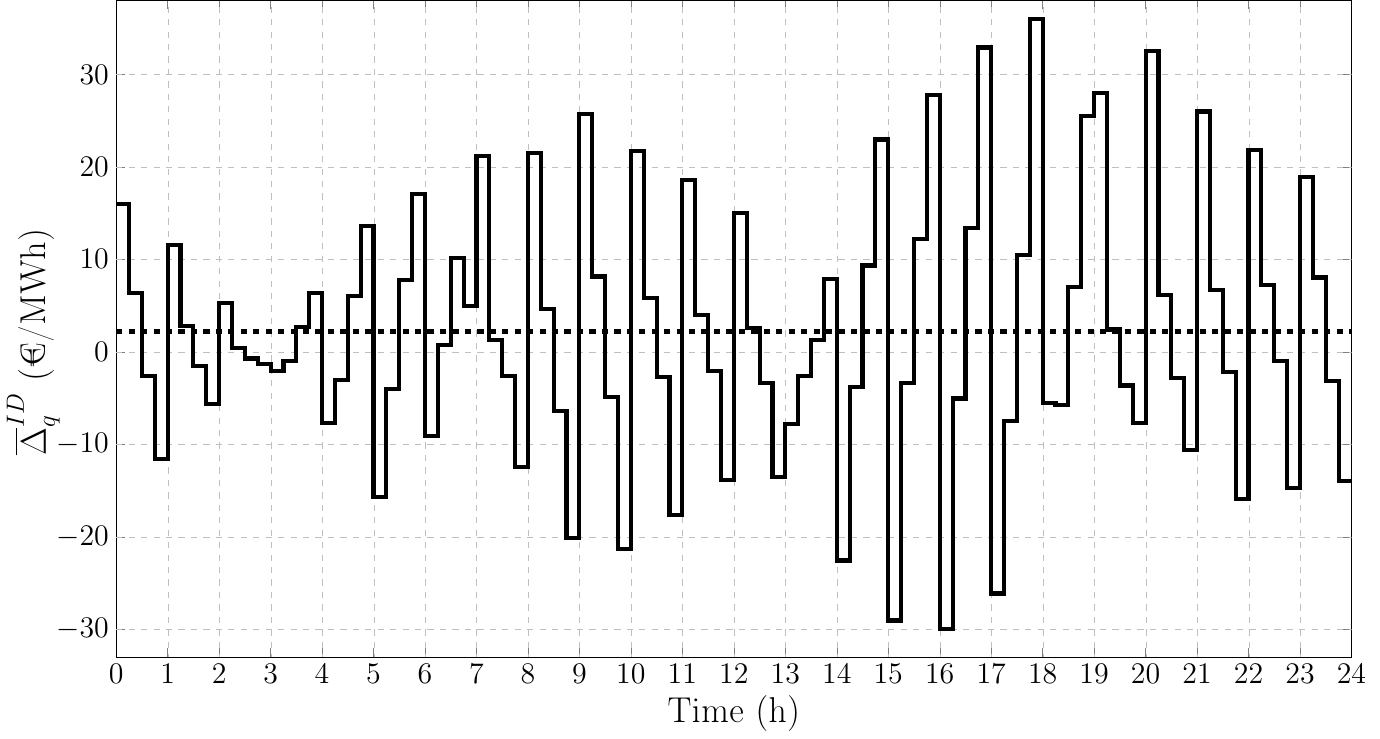}
	\caption{Average ID-DA price deviation $\overline{\Delta}_{q}^{ID}$ for the full year 2023. Mean value indicated by dashed line.}
	\label{fig:Delta_ID}
\end{figure}

\begin{figure}[t]
	\centering    
    \begin{subfigure}[b]{0.475\textwidth}
    \includegraphics[width=\linewidth]{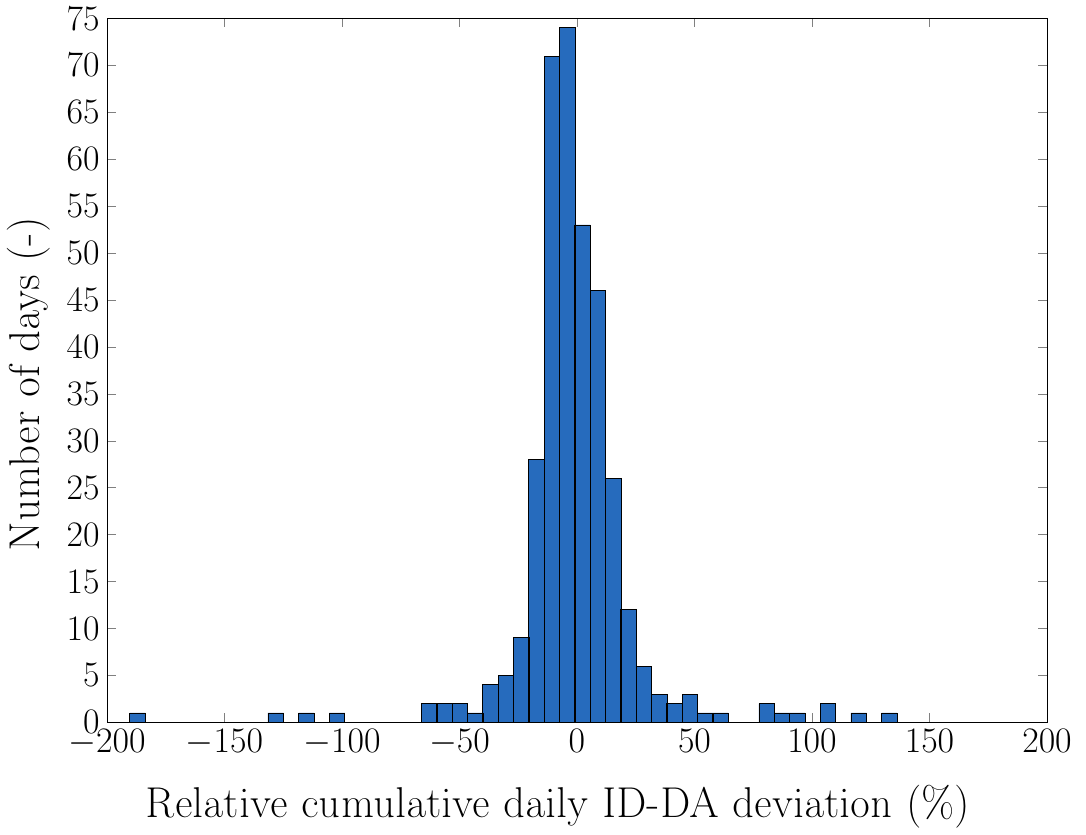}
    \caption{}
    \label{fig:integral-deviation}
    \end{subfigure}
    \hfill
    \begin{subfigure}[b]{0.49\textwidth}
    \includegraphics[width=\linewidth]{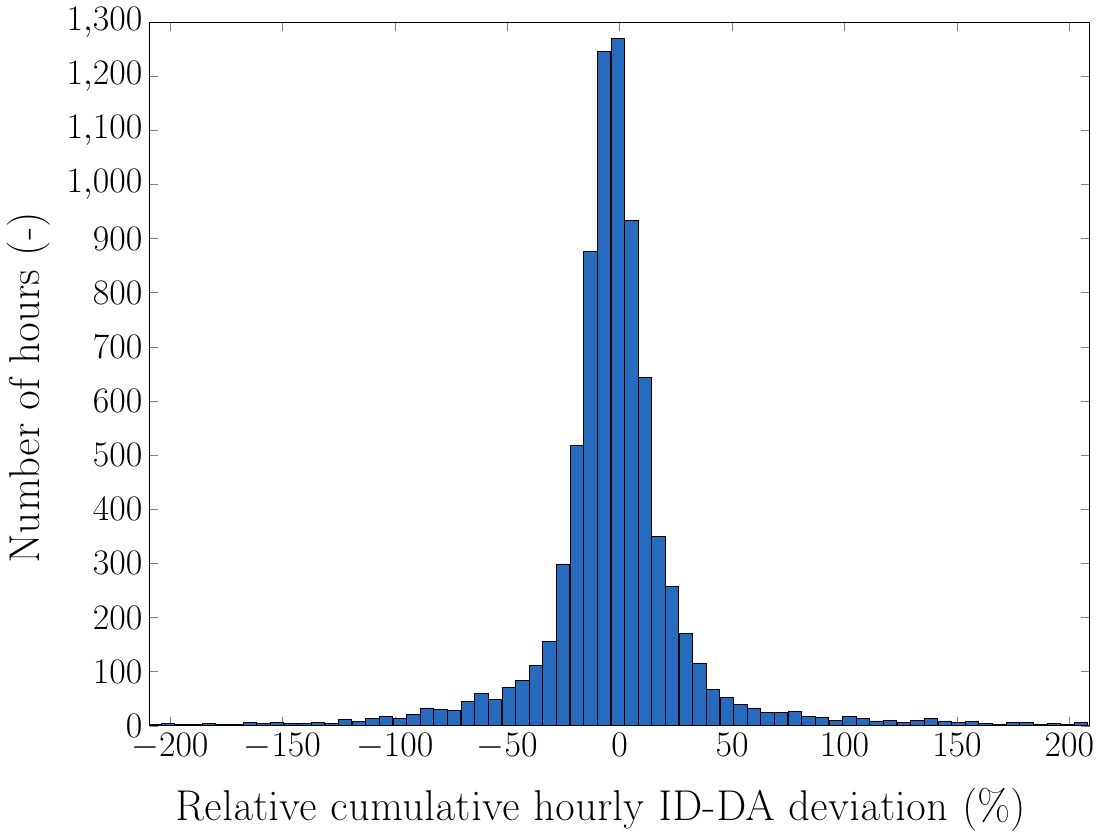}
    \caption{}
    \label{fig:integral-deviation-hour}
    \end{subfigure}    
	\caption{Distribution histogram of relative cumulative daily (a) and hourly (b) deviation of ID and DA prices. Based on historical data of year 2023.}
\end{figure}

We first calculate the average market deviation of ID around the DA profile for each \unit[15]{min} interval of the historical data ($\overline{\Delta}_{q}^{ID}$) similar to \cite{Germscheid.2022}:
\begin{equation}
    \label{eq:IDDAaver}
    \begin{split}
    \Delta_{i,q}^{ID}
    &= ID_{i,q} - DA_{i,\lceil q/4 \rceil} \,,\\
    \overline{\Delta}_{q}^{ID} 
    &= \frac{1}{N} \sum_{i=1}^{N}
    \Delta_{i,q}^{ID} \,,
    \end{split}
\end{equation}
where $q = 1, ..., 96$ is the quarter hour index,
$ID_{i,q}$ is the ID price on day $i$ and quarter hour $q$, 
and $\lceil \,\cdot\, \rceil$ denotes the ceiling operation utilized to translate from quarter hour index $q$ to corresponding hour index $k$.
Fig.~\ref{fig:Delta_ID} shows an exemplary profile of $\overline{\Delta}_{q}^{ID}$ based on the full year 2023.
Therein, we observe the deviation of an hourly pattern as indicated by a Fourier analysis \citep{Germscheid.2022}.
Notably, large jumps in the price deviation occur at full hours, typically involving a change of sign.
Moreover, we observe phase shifts at hours 3, 7, 13, and 19, where the inner hour trend switches between continuously increasing and decreasing.
Next, a straightforward approach for constructing an ID profile would be:
\begin{equation}
\label{eq:vanilla}
\overline{ID}_q = \overline{DA}_{\lceil q/4 \rceil} + 
\overline{\Delta}_{q}^{ID}
\end{equation}
However, to meet Requirement \ref{req:comarable}
we refine this approach by imposing the additional condition of zero cumulative market deviation:
\begin{equation}
    \label{eq:integral}
    \sum_{q=1}^{96} \overline{\Delta}_{q}^{ID}= 0
    \,,
\end{equation}
such that the cumulative prices of the DA and ID profiles will be identical.
Figure~\ref{fig:integral-deviation} reveals that the daily cumulative deviation scaled by the daily cumulative DA price in 2023 is approximately symmetrically distributed around zero, which supports Eq.~\eqref{eq:integral}.
Moreover, even the hourly integral of the ID to DA price deviations scaled by the hourly cumulative DA price appears to be symmetric around zero mean, Fig.~\ref{fig:integral-deviation-hour}.
This observation is consistent with the dominant harmonic frequencies at $\unit[0.5]{h^{-1}}$ and $\unit[1]{h^{-1}}$ mentioned above.
We found similar distributions for the years 2019 to 2022 (figures omitted for brevity).
The distributions, although approximately zero-mean, exhibit significant deviations.
Hence, they only valid to first order.

Because the distribution in Fig.~\ref{fig:integral-deviation} is, however, not perfectly Gaussian and exhibits non-zero mean (\unit[2.23]{\euro/MWh}, also see Fig.~\ref{fig:Delta_ID}),
Eq.~\eqref{eq:IDDAaver}
does not satisfy the consistency requirement (Eq.~\eqref{eq:integral}).
Hence, we modify the approach and extend Eq.~\eqref{eq:vanilla} by an offset correction term to obtain:
\begin{equation}
    \label{eq:IDDA-corr}
    \tilde{\Delta}_{q}^{ID}
    = \overline{\Delta}_{q}^{ID} 
    - \frac{1}{96} \sum_{q=1}^{96}
    \overline{\Delta}_{q}^{ID} \,.
\end{equation}
We notice that the correction approach may be further refined based on the zero-mean distribution in Fig.~\ref{fig:integral-deviation-hour}.
However, we decide against such an additional refinement to keep the method simple. 
We remark that analysis of the years 2019 to 2022 revealed both positive and negative correction terms of smaller orders of magnitude compared to 2023 (\unit[2.23]{\euro/MWh}).
This indicates that the correction assumption is generally valid and introduces minimal changes to the averaged historical data.
Next, similar to $\beta$ for the DA profile, we scale $\tilde{\Delta}^{ID}_q$ by a parameter $\gamma>0$ to match the year-average of the ID standard deviations:
\begin{equation}\label{eq:gammastd}
  \sigma(\overline{ID}_1^\gamma ..., \overline{ID}_{96}^\gamma)
  = \frac{1}{N} \sum_{i=1}^{N} 
  \sigma({ID}_{i,1}, ..., {ID}_{i,96}) \,,
\end{equation}
wherein the constructed ID profile is calculated as:
\begin{linenomath}\begin{equation*}
    \overline{ID}_q^\gamma = \overline{DA}_{\lceil q/4 \rceil}^\beta + 
    \gamma \tilde{\Delta}_{q}^{ID} \,,\, q=1,...,96 \,.
\end{equation*}\end{linenomath}
The parameter value $\gamma$ corresponding to the \textit{Nominal} profile is specified by solving the implicit nonlinear equation with $\beta$ fixed:
\begin{equation}
    \label{eq:scalinggamma}
	\sqrt{
    \sum_{q=1}^{96}(\beta (\overline{DA}_{\lceil q/4 \rceil} - \overline{DA}) + \gamma \tilde{\Delta}_{q}^{ID})^2}
    =
    \frac{1}{N}
    \sum_{i=1}^{N}
    \sqrt{
    \sum_{q=1}^{96}
    \big({ID}_{i,q} - 
    \frac{1}{96}\sum_{q=1}^{96} {ID}_{i,q}
    \big)^2
    }
    \,.
\end{equation}
Equation \eqref{eq:scalinggamma} is readily solved numerically, e.g., using 
Microsoft Excel.
For the full year 2023, we determine $\gamma = 1.91$.

Again, we may adjust the fluctuation magnitude of an ID profile around its DA reference through a modification of $\gamma$.
For example, when specifying $\gamma$ such that the standard deviation matches the quantile 3.4 of the historical ID standard deviation density distribution:
\begin{equation}
\begin{split}
    \label{eq:scalinggammaquantile}
    Q_{3.4}\big(\{\sigma(ID_{i,1}, ..., ID_{i,96})\}_{i=1}^N\big) 
   &= \sigma(\overline{ID}_{1}^{\gamma}, ..., \overline{ID}_{96}^{\gamma}) \\
   &=
    \sqrt{
    \frac{1}{96}
    \sum_{q=1}^{96}(\beta (\overline{DA}_{\lceil q/4 \rceil} - \overline{DA}) + 
    \gamma \tilde{\Delta}_{q}^{ID})^2} \,,
    \end{split}
\end{equation}
we obtain $\gamma = 1.77$ for 2023. We here note that although the $\gamma$-value in the \textit{Extreme} scenario is smaller than that of the \textit{Nominal} scenario, the overall deviation in the \textit{Extreme} ID profile is higher, due to the higher $\beta$-value of the \textit{Extreme} DA profile taking part in the calculations (Eq.~\eqref{eq:scalinggammaquantile}).
Similar to the DA prices, larger and smaller values of $\gamma$ create more aggressive and less aggressive ID profile, respectively.

\begin{figure}[t]
	\centering    
    \includegraphics[width=0.8\linewidth]{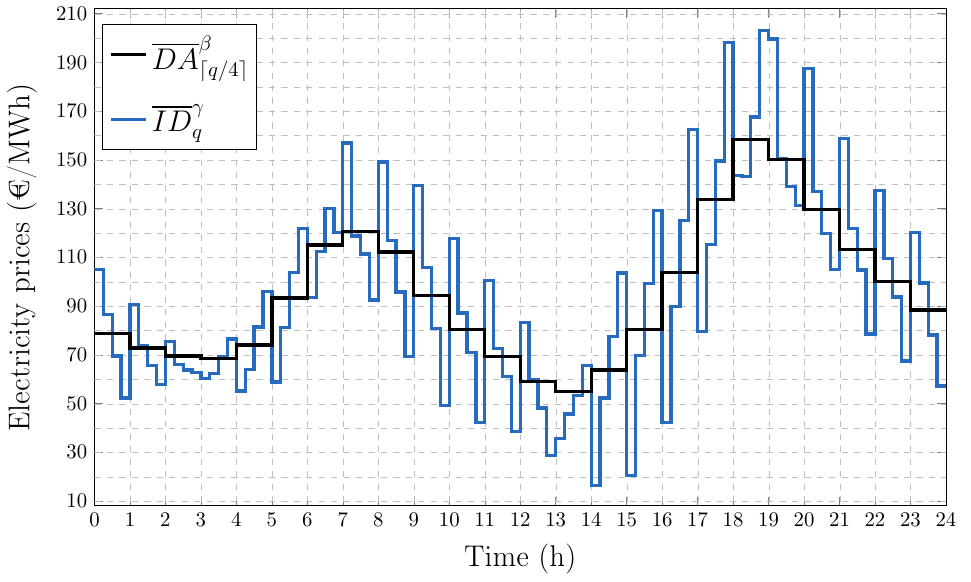}
	\caption{\textit{Nominal} ($\beta=1.47$, $\gamma=1.91$) single-day DA and ID price profile for year 2023 \citep{EPEX}.}
	\label{fig:BenchmarkProfile}
\end{figure}

Figure~\ref{fig:BenchmarkProfile} illustrates the final single-day \textit{Nominal} ready-to-use DA and ID profile based on full year 2023 data and for $\beta = 1.47$ and $\gamma = 1.91$.
For the DA profile we note that prices exhibit a daily minimum, coinciding with peak photovoltaic generation around noon \citep{Schafer.2020, Kiesel.2017}. 
Prices start rising around 6:00, when the workday begins and start falling around 18:00 when the workday ends. 
Two price peaks around 7:00 and 18:00 denote a high electricity demand meeting a low renewable energy supply from photovoltaics \citep{Knittel.2005}.
On the other hand, ID prices drop subhourly between 07:00 to 13:00 and 19:00 to 03:00 and rise in the meantime.
Namely, while the sun is rising (from 07:00 to 13:00), offer is higher than the hourly average demand and thus price is lower in the last quarter of the hour.
During afternoon hours (from 13:00 to 19:00), the offer is higher in the first quarter of the hour. 
Further, during offpeak hours (from 19:00 to 07:00) the oscillation of the average ID price follows the demand of power-intensive industry (decreasing steps), and the production design of the fossil power plants (increasing steps) for minimum production regulations \citep{Kiesel.2017}.

\newpage
\vspace{-1ex}
\section{Single-week price profile}\label{sec:OneWeekPriceConstruction}\label{sec:priceprofileS1723_week} 
\vspace{-1ex}
In this section, we adust the methodology presented in the previous sections to generate single-week DA (Section \ref{sec:DAconstructionWeek}) and ID (Section \ref{sec:IDconstructionWeek}) price profiles from the full-year time-series data set.

\subsection{DA price profile}
\label{sec:DAconstructionWeek} 
\vspace{-1ex}
Commonly, the time horizon of planning and scheduling spans more than one day of operation.
In some cases, a periodic scenario is sufficient, wherein the price profile is duplicated for multiple consecutive days.
However, if distinct prices are desired, the procedure from the previous section can be extended to price scenarios comprising multiple days.
Here, we construct such a price profile for a one-week time window.
The choice of this horizon length aligns with the significant DA price frequencies pinpointed by \cite{Germscheid.2022}, allowing us to capture and average over weekly patterns. 
However, we do not recommend to apply the averaging method to construct time frames longer than a week.
In that case, the averaging would involve either statistically too small historical data sets (for $N=365$) or data sets spanning more than a year and thus including long-term changes of market dynamics and policy.
However, one remedy to construct a profile spanning multiple weeks could be the periodic repetition of the averaged profile.

Following the approach presented in Section \ref{sec:DAconstruction}, we first average the historical DA prices for every hour $24d+k = 1,2,...,168$ of the week after splitting the available data in an integer number of full weeks $W$ (Monday to Sunday), so that $7\cdot W\leq N$:
\begin{linenomath}\begin{equation*}
    \label{eq:averageDA_week}
	\overline{DA}_{24d+k} = \frac{1}{W} 
    \sum_{w=1}^{W}{DA_{(d+1)+7(w-1),k}}
    \,,
\end{equation*}\end{linenomath}
where $d = 0,1,...,6$ indicates the day of the week $w = 1,...,W$ (e.g, Monday corresponds to 0), $k = 1,2,...,24 $ denotes the hour of the day and $24d+k$ is the hour of the constructed single-week profile. Similar to Section \ref{sec:DAconstruction} we use the hourly data of 2023, thus considering $W=52$.

We next calculate the overall mean value:
\begin{linenomath}\begin{equation*}
\label{eq:meanofDA_week}
    \overline{DA} = \frac{1}{168} 
    \sum_{k=1}^{24}
    \sum_{d=0}^{6}
    \overline{DA}_{24d+k} \,,
\end{equation*}\end{linenomath}
and scale the profile as in Eq.~\eqref{eq:DApricescaling} by factor $\beta$:
\begin{linenomath}\begin{equation*}
    \label{eq:DApricescaling_week}
	\overline{DA}_{24d+k}^{\beta} = \overline{DA} + \beta \cdot (\overline{DA}_{24d+k} - \overline{DA})
    \,.
\end{equation*}\end{linenomath}
For the construction of a \textit{Nominal} profile, $\beta$ is selected such that the standard deviation of the constructed single-week profile is equal to the average standard deviation of all weekly profiles of the historical data:
\begin{linenomath}\begin{equation*}
  \sigma(\overline{DA}_{1}^{\beta}, ..., \overline{DA}_{168}^{\beta}) 
  = \frac{1}{W} \sum_{w=1}^{W} \sigma(DA_{7w-6,1}, ..., DA_{7w-6,24}, DA_{7w-5,1}, ..., DA_{7w,24}) \,,
\end{equation*}\end{linenomath}
which leads to the explicit $\beta$ expression:
\begin{equation}
    \label{eq:scalingbeta_week}
	\beta =
    \frac{1}{W}
    \sum_{w=1}^{W}
    \sqrt{
    \frac{
    \sum_{d=0}^{6}
    \sum_{k=1}^{24}
    \big( DA_{(d+1)+7(w-1),k} - 
    \frac{1}{168}
    \sum_{d=0}^{6}
    \sum_{k=1}^{24}
    DA_{(d+1)+7(w-1),k}
    \big)^2
    }{
    \sum_{d=0}^{6}
    \sum_{k=1}^{24}(
    \overline{DA}_{24d+k} - \overline{DA}
    )^2
    }}
    \,.
\end{equation}
For more extreme price fluctuations, we select $\beta$ so that the standard deviation of the constructed profile is equal to the quantile 3.0 value of the density distribution:
\begin{equation}
    \label{eq:scalingbetaextreme_week}
	\beta =
    \frac{Q_{3.0}\big( \{ \sigma(DA_{7w-6,1}, ..., DA_{7w-6,24}, DA_{7w-5,1}, ..., DA_{7w,24})\}_{w=1}^W\big)}{
    \sqrt{
    \frac{1}{168}
    \sum_{d=0}^{6}
    \sum_{k=1}^{24}(
    \overline{DA}_{24d+k} - \overline{DA}
    )^2
    }}
    \,.
\end{equation}
We here use a smaller quantile compared to the one selected in Section~\ref{sec:DAconstruction}, because weeks with extremely varying prices do not usually follow the nominal fluctuating pattern, see also figures of the supplementary information (SI).

Considering the EPEX Spot DA market data for the year 2023, Eq.~\eqref{eq:scalingbeta_week} gives $\beta=1.58$, and Eq.~\eqref{eq:scalingbetaextreme_week} $\beta=1.76$. 
More or less aggressive profiles are again obtained by adjusting the parameter $\beta$.

\subsection{ID price profile}\label{sec:IDconstructionWeek}
\vspace{-1ex}
We extend our methodology on extracting a practical ID profile directly following the analysis in Section \ref{sec:IDconstruction} to compose the ID profile over a one-week horizon by superposing the average ID-DA price deviation with the average DA price over a single-week period. Therefore, we first calculate the average market deviation of the historical data:
\begin{linenomath}\begin{equation*}
    \label{eq:IDDAaver_week}
    \begin{split}
    \overline{\Delta}_{96d+q}^{ID}
    &= \frac{1}{W} \sum_{w=1}^{W}
    \big(ID_{(d+1)+7(w-1),q} - DA_{(d+1)+7(w-1),\lceil q/4 \rceil}\big) \,,\\
    \end{split}
\end{equation*}\end{linenomath}
where $q = 1, ..., 96$ is the quarter hour index,
$ID_{(d+1)+7(w-1),q}$ is the ID price on week $w$, day of the week $d$ and quarter hour $q$.

As opposed to the hourly and daily historical profiles, the weekly profiles do not show a zero-mean distribution of the weekly cumulative ID-DA deviation.
Thus, a zero-mean correction for the full week as in Eq.~\eqref{eq:IDDA-corr} is not expedient here.
Instead, we apply Eq.~\eqref{eq:IDDA-corr} separately for each of the seven days of the week to enforce a daily closure:
\begin{linenomath}\begin{equation*}
    \label{eq:integral_week}
    \sum_{q=1}^{96}
    \overline{\Delta}_{96d+q}^{ID}
    = 0 
    \,,\hspace{3ex} d=0,...,6
\end{equation*}\end{linenomath}
and therefore:
\begin{linenomath}\begin{equation*}
    \label{eq:IDDA-corr_week}   \tilde{\Delta}_{96d+q}^{ID}
    = \overline{\Delta}_ {96d+q}^{ID}
    - \frac{1}{96} \sum_{q=1}^{96}
    \overline{\Delta}_{96d+q}^{ID} \,,\hspace{3ex} d=0,...,6\,.
\end{equation*}\end{linenomath}
As a result, the single-week profile fulfills Requirement~\ref{req:comarable}.

\begin{figure}[ht]
	\centering    
    \includegraphics[width=0.98\linewidth]{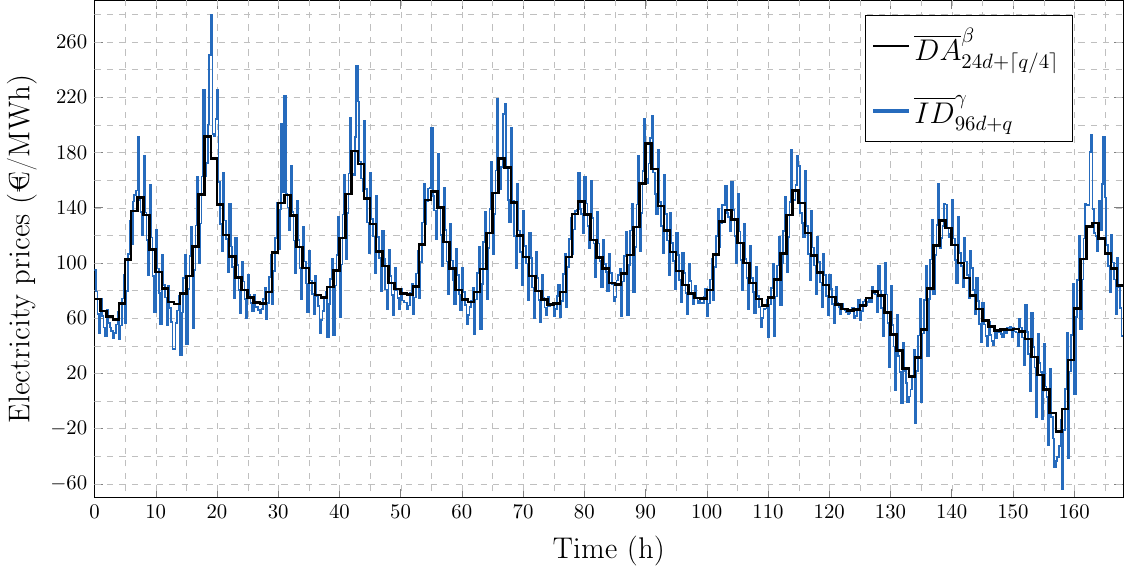}
	\caption{\textit{Nominal} ($\beta=1.58$, $\gamma=1.60$) single-week DA and ID price profile for year 2023.}
	\label{fig:BenchmarkProfileNominalWeek}
\end{figure}

Again, the final ID profile is constructed by the superposition of the $\gamma$ scaled market deviations to the single-week constructed DA profile:
\begin{linenomath}\begin{equation*}
    \overline{ID}_{96d+q}^\gamma = \overline{DA}_{24d + \lceil q/4 \rceil}^\beta + 
    \gamma \tilde{\Delta}_{96d+q}^{ID} \,,\hspace{3ex} d=0,...,6 \,,\, q=1,...,96 \,.
\end{equation*}\end{linenomath}
As in Eq.~\eqref{eq:gammastd}, we select $\gamma$ by solving the equation:
\begin{equation}\label{eq:gammastd_week}
\begin{split}
  \frac{1}{W} \sum_{w=1}^{W} \sigma(ID_{7w-6,1}, ..., ID_{7w-6,96}, &ID_{7w-5,1}, ..., ID_{7w,96}) 
  \\
  &=\sqrt{
    \sum_{d=0}^{6}
    \sum_{q=1}^{96}\big(\beta (\overline{DA}_{24d+\lceil q/4 \rceil} - \overline{DA}) + \gamma \tilde{\Delta}_{96d+q}^{ID}\big)^2}
    \end{split}
\end{equation}
for $\gamma>0$.
In this way, the constructed single-week ID profile reproduces the standard deviation of the averaged standard deviation of the historical single-week ID profiles.
For a stronger higher varying profile, the parameter $\gamma$ can be calculated as:
\begin{equation}
\begin{split}
    \label{eq:scalinggammaquantile_week}
    Q_{3.0}\big(\{\sigma(ID_{7w-6,1}, ..., ID_{7w,96})\}_{w=1}^{W}\big) 
    =
  \sqrt{
    \sum_{d=0}^{6}
    \sum_{q=1}^{96}\big(\beta (\overline{DA}_{24d+\lceil q/4 \rceil} - \overline{DA}) + \gamma \tilde{\Delta}_{96d+q}^{ID}\big)^2}
    \,.
    \end{split}
\end{equation}
Considering the historical data of EPEX Spot for 2023, and 52 full weeks (from Monday to Sunday), Eq.~\eqref{eq:gammastd_week} gives $\gamma = 1.60$, and Eq.~\eqref{eq:scalinggammaquantile_week} gives $\gamma = 2.32$. The final single-week \textit{Nominal} ready-to-use DA and ID profiles for the full year 2023 data ($\beta = 1.58$,$\gamma = 1.60$) are shown in Fig.~\ref{fig:BenchmarkProfileNominalWeek}.

Figure~\ref{fig:BenchmarkProfileNominalWeek} is consistent with the observations of \cite{Germscheid.2023}. 
Lowest prices are reached during the weekend associated with a lower demand.  
Accordingly, the price peak in the weekend mornings is less significant than the peak in the evening.  
As reported by \cite{Guthrie.2002}, during weekdays the daytime off-peak and evening peak periods exhibit distinct market behaviors.
However, during the weekend these periods operate as a unified market.
Additionally, the absence of morning peaks in the weekends results from morning prices correlated with the rest prices of the weekend.
The highest DA prices are reached on Monday and Thursday evening, and the highest ID prices on Monday evening.

To capture seasonal price patterns, the approach could be extended by averaging over a 3-month period. 
However, we consider the data from a single calendar year too limited for this technique.
At the same time, using data from 2022 and 2023 is not feasible due to significant deviations caused by the 2022 energy crisis.


\section{Exemplary price profiles}\label{sec:Results}
\vspace{-1ex}
In this section, we present the price scenarios derived from our methodology application (Section \ref{sec:profileS1723}) and compared the resulting profiles to historical time-series (Section \ref{sec:comphistorical}).

\subsection{Scenarios}
\label{sec:profileS1723}

\vspace{-1ex}
We apply the proposed method to calculate DA and ID price profiles based on the full year 2023 historical data (01.01.2023 to 31.12.2023).
We construct profiles following Sections \ref{sec:OneDayPriceConstruction} and \ref{sec:OneWeekPriceConstruction} with three variance scaling cases each: 
\begin{enumerate}
    \vspace{-2ex}
    \setlength\itemsep{0ex}
    \item Single-day horizon: \newline
    i) \emph{Nominal} ($\beta=1.47$, $\gamma=1.91$), ii) \emph{Unscaled} ($\beta=1.0$, $\gamma=1.0$), iii) \emph{Extreme} ($\beta=1.85$, $\gamma=1.77$),
    \item Single-week horizon: \newline
    i) \emph{Nominal} ($\beta=1.58$, $\gamma=1.6$), ii) \emph{Unscaled} ($\beta=1.0$, $\gamma=1.0$), iii) \emph{Extreme} ($\beta=1.76$, $\gamma=2.32$).
\end{enumerate}

We show the resulting single-day \textit{Nominal} profiles in Fig.~\ref{fig:BenchmarkProfile} and provide the \textit{Unscaled} and \textit{Extreme} profile in the SI. 
Similarly, the single-week \textit{Nominal} profiles are shown in Fig.~\ref{fig:BenchmarkProfileNominalWeek} and the \textit{Unscaled} and \textit{Extreme} profiles can be found in the SI.  
We furthermore provide the profiles in \texttt{CSV} format via Git (\git) as well as in table format in the SI. 
We highlight that the \emph{Unscaled} data may be used as a basis to construct price profiles for arbitrary scaling options ($\beta$, $\gamma$).

\begin{figure}[hpt]
	\centering    
 \begin{subfigure}[b]{0.99\textwidth}
         \includegraphics[width=0.95\linewidth]{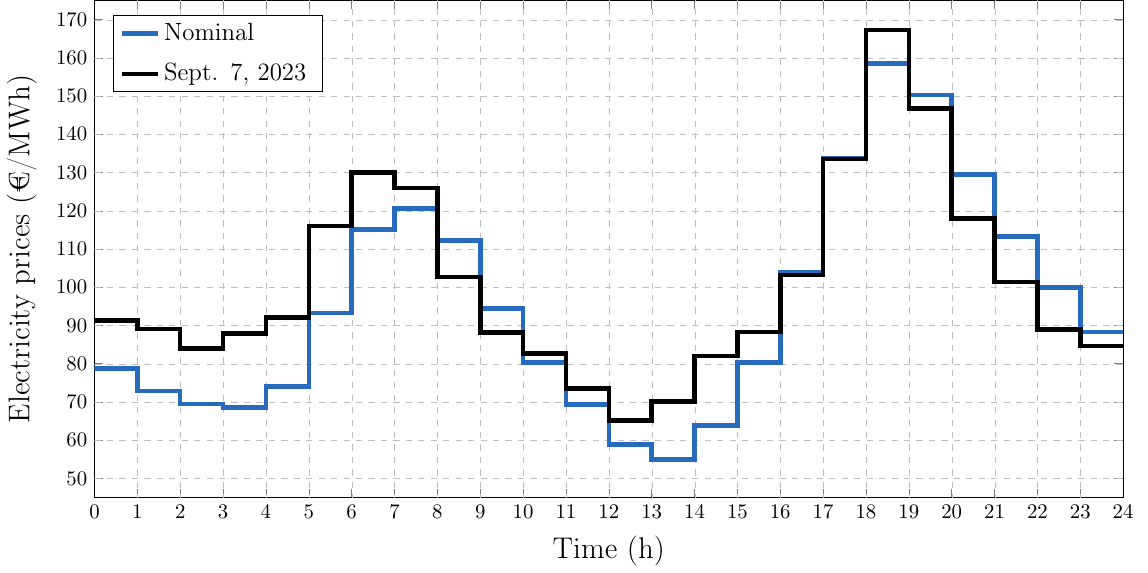}
     \caption{DA profiles}\vspace{5ex}
	\label{fig:HistoricDayDANominal
 }
 \end{subfigure}
 \begin{subfigure}[b]{0.99\textwidth}
         \includegraphics[width=0.95\linewidth]{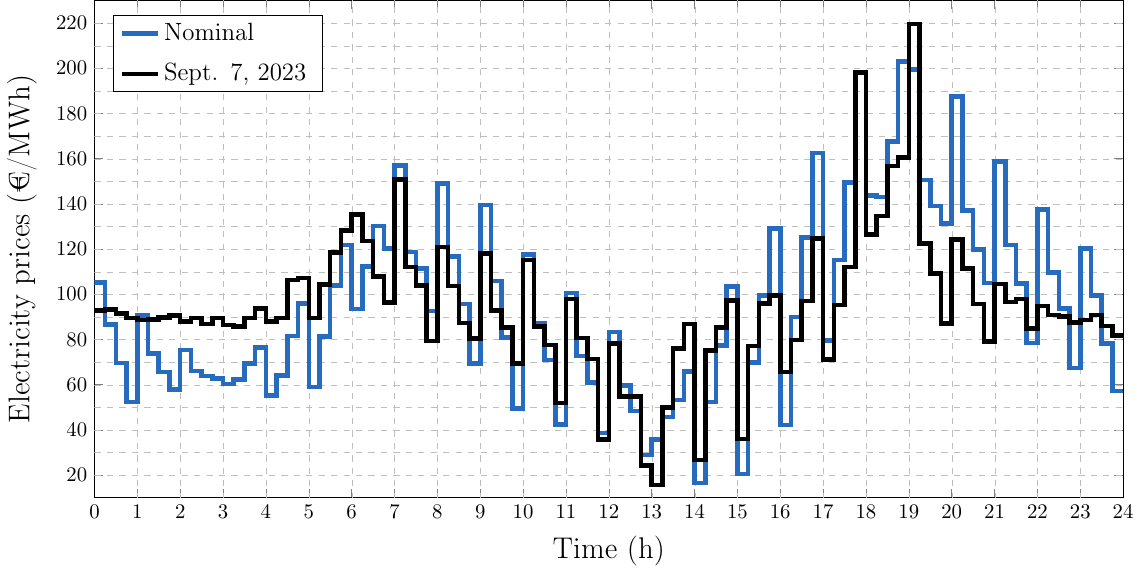}
    \caption{ID profiles}
	\label{fig:HistoricDayIDNominal}
 \end{subfigure}
 	\caption{\textit{Nominal} single-day and corresponding best fit of historical profiles for year 2023.}
  \label{fig:HistoricDayDA}
\end{figure}

\begin{figure}[hpt]
	\centering    
 \begin{subfigure}[b]{0.99\textwidth}
    \includegraphics[width=0.95\linewidth]{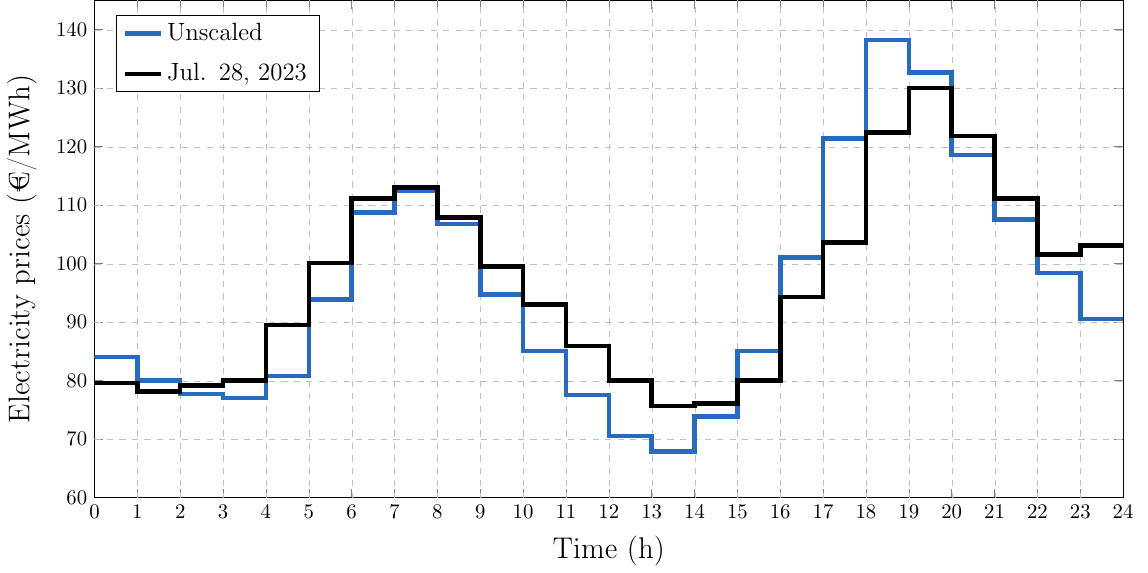}
	\caption{DA profile}\vspace{5ex}
	\label{fig:HistoricDayDAModerate}
\end{subfigure}
 \begin{subfigure}[b]{0.99\textwidth}
	\centering    
    \includegraphics[width=0.95\linewidth]{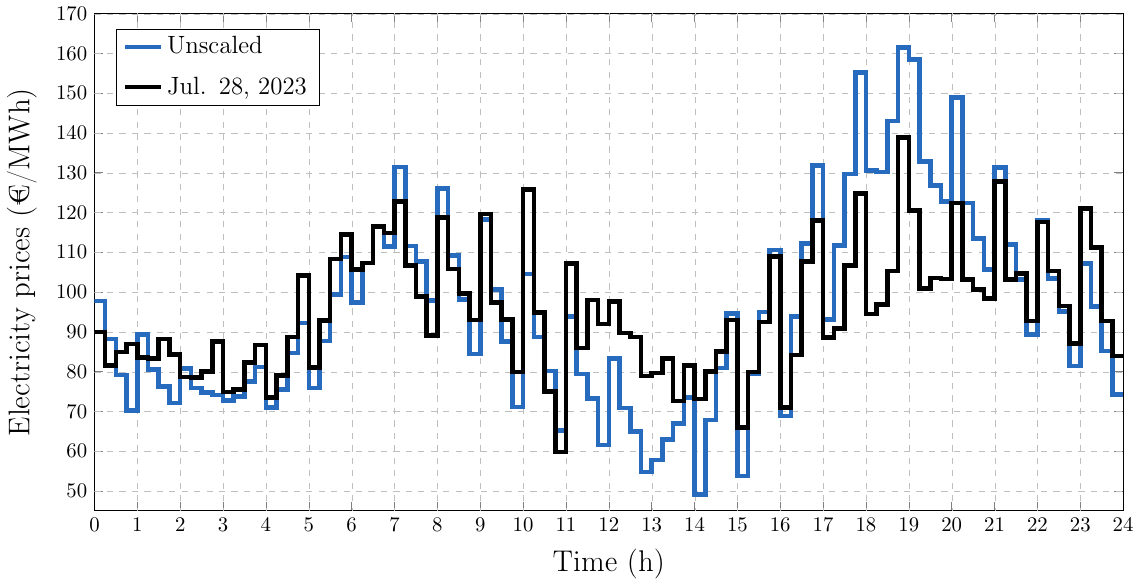}
	\caption{ID profile}
	\label{fig:HistoricDayIDModerate}
 \end{subfigure}
 	\caption{\textit{Unscaled} single-day and corresponding best fit of historical profile for year 2023.}
 \label{fig:HistoricDayModerate}
\end{figure}

 \begin{figure}[hpt]
	\centering  
    \includegraphics[width=0.96\linewidth]{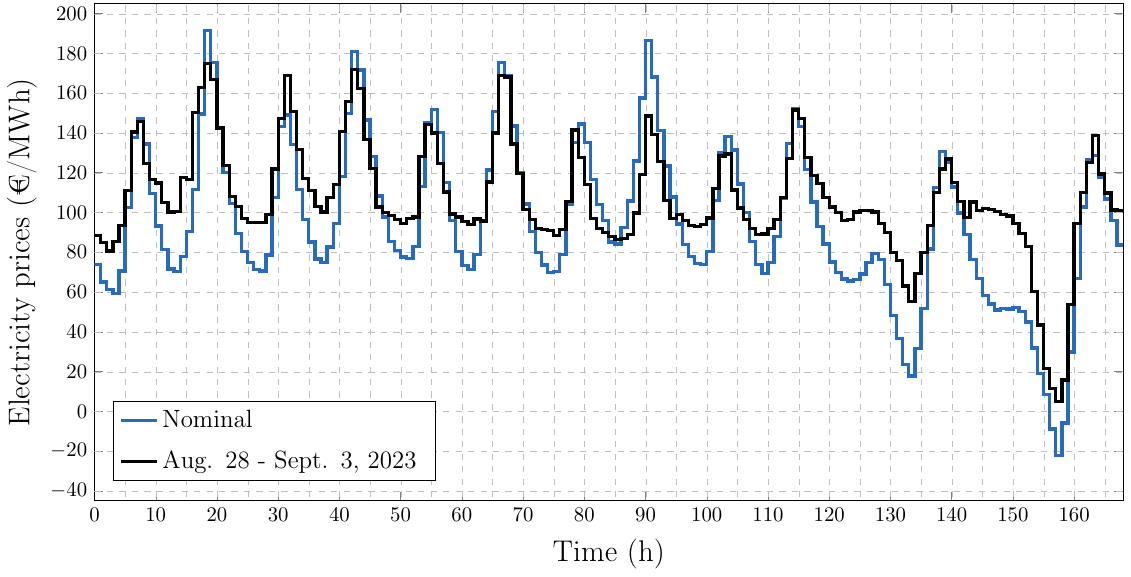}
	\label{fig:HistoricWeekDA}
 	\caption{\textit{Nominal} DA single-week price profiles compared to the best fit historical week in 2023.}
 \label{fig:HistoricWeek}
\end{figure}

We additionally provide a quantitative statistical evaluation of the constructed profiles in Tables \ref{tab:StatDA} to \ref{tab:StatID_week}.
In particular, the tables confirm that
the standard deviation of the \textit{Nominal} DA and ID profiles and the corresponding historical data match as intended by nominal scaling.
Moreover, while the mean of the DA profiles matches the historical data,
there is a difference for the ID profiles due to the correction of the integral zero-mean deviation.  
Consequently, the \unit[24]{h} price integrals of the corresponding DA and ID profiles are identical (compare last row in Tables \ref{tab:StatDA} and \ref{tab:StatID} as well as Tables \ref{tab:StatDA_week} and \ref{tab:StatID_week}).

\begin{table}[p]
\begin{minipage}{\textwidth}
\centering
    \caption{Characteristic statistical values of the single-day DA electricity price profile.}
    \label{tab:StatDA}
\begin{tabular}{l rrrr}
    \multicolumn{5}{l}{\footnotesize $^\ast$ Average value over all week integrals}
    \\[0ex]\multicolumn{5}{l}{\footnotesize $^\bigstar$ The number of histogram bins is calculated with the Scott rule \citep{Scott.1979}}
    \\[0ex]\multicolumn{5}{l}{\footnotesize $^\triangledown$ Average value over all full weeks}
    \\[2ex]
    \toprule
    \textbf{Quantity} & \multicolumn{4}{c}{\textbf{DA price profile}}
    \\ 
     & \textit{Nominal} &\textit{Unscaled} & \textit{Extreme} & historical 
    \\ 
    \midrule
    $\beta$ & 1.47 & 1.00 & 1.85$^\bigstar$ & - \\
    Minimum (\euro/MWh) & 54.99 & 67.82 & 67.82 & $-500.00$ \\ 
    Maximum (\euro/MWh) & 158.45 & 138.24 & 138.24 & 524.27 \\ 
    Mean (\euro/MWh) & 95.18 & 95.18 & 95.18 & 95.18\\ 
    Std.~deviation (\euro/MWh) & 28.22 & 19.21 & 35.46 & 28.22
    \\ 
    \unit[24]{h} integral (\euro/24MWh) & 2284.21 & 2284.21 & 2284.21 & 2284.21$^\ast$ \\ 
    \bottomrule
\end{tabular}
\vspace{3ex}
\centering
    \caption{Characteristic statistical values of the single-day ID electricity price profile.}
    \label{tab:StatID}
\begin{tabular}{l rrrr}
    \toprule
    \textbf{Quantity} & 
    \multicolumn{4}{c}{\textbf{ID price profile}}
    \\ 
     & \textit{Nominal} & \textit{Unscaled} & \textit{Extreme} & historical 
    \\ 
    \midrule
    $\gamma$ & 1.91 & 1.00 & 1.77 & - \\
    Minimum (\euro/MWh) & 16.36 & 49.04 & 12.00 & -666.87 \\ 
    Maximum (\euro/MWh) & 202.96 & 161.50 & 215.73 & 3543.51 \\ 
    Mean (\euro/MWh) & 95.18 & 95.18 & 95.18 & 97.40 \\ 
    Std.~deviation (\euro/MWh) & 40.31 & 24.61 & 44.73 & 40.31
    \\ 
    \unit[24]{h} integral (\euro/24MWh) & 2284.21 & 2284.21 & 2284.21 & 2337.64$^\ast$ \\ 
    \bottomrule
\end{tabular}
\vspace{3ex}
\centering
    \caption{Characteristic statistical values of the single-week DA electricity price profile.}
    \label{tab:StatDA_week}
\begin{tabular}{l rrrrr}
    \toprule
    \textbf{Quantity} & \multicolumn{4}{c}{\textbf{DA price profile}}
    \\ 
     & \textit{Nominal} & \textit{Unscaled} & \textit{Extreme} & historical 
    \\ 
    \midrule
    $\beta$ & 1.58 & 1.00 & 1.76$^\bigstar$ & - \\
    Minimum (\euro/MWh) & -22.07 & 20.92 & -35.70 & -500.00 \\ 
    Maximum (\euro/MWh) & 191.34 & 156.22 & 202.47 & 524.27 \\ 
    Mean (\euro/MWh) & 95.40 & 95.40 & 95.40 & 95.40$^\triangledown$ \\ 
    Std.~deviation (\euro/MWh) & 38.59 & 24.46 & 43.07 & 38.59$^\triangledown$
    \\ 
    \unit[1]{week} integral (\euro/24MWh) & 16026.52 & 16026.52 & 16026.52 & 16026.52$^\ast$ \\ 
    \bottomrule
\end{tabular}
\vspace{3ex}
\centering
    \caption{Characteristic statistical values of the single-week ID electricity price profile.}
    \label{tab:StatID_week}
\begin{tabular}{l rrrr}
    \toprule
    \textbf{Quantity} & \multicolumn{4}{c}{\textbf{ID price profile}}
    \\ 
     & \textit{Nominal} & \textit{Unscaled} & \textit{Extreme} & historical 
    \\ 
    \midrule
    $\gamma$ & 1.60 & 1.00 & 2.32$^\bigstar$ & -\\
    Minimum (\euro/MWh) & -64.06 & -5.18 & -101.93 & -666.87 \\ 
    Maximum (\euro/MWh) & 279.91 & 211.45 & 336.05 & 3543.51 \\ 
    Mean (\euro/MWh) & 95.40 & 95.40 & 95.40 & 95.61$^\triangledown$ \\ 
    Std.~deviation (\euro/MWh) & 47.02 & 29.69 & 57.48 & 47.02$^\triangledown$ 
    \\ 
    \unit[1]{week} integral (\euro/24MWh) & 16026.52 & 16026.52 & 16026.52 & 16026.52$^\ast$ \\ 
    \bottomrule
\end{tabular}
\end{minipage}
\end{table}

\subsection{Comparison to historical data}\label{sec:comphistorical}
\vspace{-1ex}

For the sake of comparison, we complement the constructed profiles by historical profiles that closely match each of the calculated profiles.
Thereby, we show that the constructed profiles do not only artificially represent the averaged historical data but are even similar to real historical price profiles.
Such a historic profile may thus be considered close to the center point in the averaging metric.
We retrieve the corresponding historical day by minimizing the mean absolute deviation between the historical and a constructed scenario for DA and ID simultaneously.
The resulting profiles are visually juxtaposed to the constructed ones.
In the single-day case, we solve:
\begin{equation}
\label{eq:minDay}
    \min_{i}
    \Big(
    \sum_{k=1}^{24}
    |\overline{DA}_{k}^{\beta} - 
    DA_{i,k}|
    +
    \frac{1}{4}
    \sum_{q=1}^{96}
    |\overline{ID}_{q}^\gamma - 
    ID_{i,q}| 
    \Big)\,.
\end{equation}

A single-week historical profile is identified analogously.
The minimization problem is solved by first calculating the absolute price differences between the constructed daily profiles and each of the historical DA and ID profiles, as given in Eq. (\ref{eq:minDay}).
Then, the objective value is calculated for each of the historical days, and the minimum value is selected.
The corresponding day represents the historical time-series data that best aligns with the constructed DA and ID profiles simultaneously.

Figure \ref{fig:HistoricDayDA} contrasts the \textit{Nominal} single-day profiles and the historical day with the best fit. 
While the inner hour variance of the \textit{Nominal} ID profile is more pronounced during early and late hours, all profiles follow similar trends and are overall comparable.
Moreover, inserting Eq.~\eqref{eq:DApricescaling} into Eq.~\eqref{eq:minDay} reveals that the historic profile in Fig.~\ref{fig:HistoricDayDA}
indeed reflects the day which deviates the least from the rest historical days, i.e., the cluster center in this sense.
Next, Fig.~\ref{fig:HistoricDayModerate} depicts the \textit{Unscaled} profiles and the historical best fit.
Since here all single-week profiles have the same historical week as the best fit, we only show the plot for the DA \textit{Nominal} profile in Fig.~\ref{fig:HistoricWeek}.
The rest of the figures for profile matching with historical data are given in the SI.

We raise attention to the higher mismatch of the two profiles in Fig.~\ref{fig:HistoricWeek} for the last hours of the week and we note that during the weekdays (from \unit[0]{h} up to \unit[120]{h}) the price pattern is well-defined due to a consistent daily cycle in electricity demand (i.e., prices rise when the workday begins, peak in the afternoon, and decline in the evening as demand shifts to residential use) \citep{Knittel.2005}. 
This does not apply to the weekend (from \unit[120]{h} up to \unit[168]{h}) when demand is more unpredictable.
We conclude that the \textit{Unscaled} and \textit{Nominal} profile represent the historical data closely.
For the \textit{Nominal} profile, the fitted historical day merely serves for validation, whereas for the \textit{Unscaled} profile the historical day corresponds to the most representative day in the historical data set.

Lastly, we discuss a limitation of our approach regarding more extreme scenarios.
Different from the close alignment of the \textit{Unscaled} and \textit{Nominal} profiles with historical data, solving the fitting problem in Eq.~\eqref{eq:minDay} for the \textit{Extreme} profile did not lead to satisfactory results.
This is explained by the extreme deviations, i.e., outliers, observed in the historical data, especially in the ID price.
These outliers, despite showing high variance, when composing the \textit{Extreme} profiles, exhibit different patterns than the averaged ones of the constructed (scaled) profiles.
Nevertheless, when fitting historical DA and ID profiles separately rather than jointly, we were able to find similar historical profiles (see figures in SI).
We conclude that while a more extreme variance scaling is generally valid,
an alternative scenario generation method should be consulted if extreme profiles with abnormal patterns are sought.

\section{Case study and method comparison}
\label{sec:compSOTA}
We apply our scenario generation method to a DSM scheduling case study and compare to scheduling results when employing price scenarios obtained using literature methods.
As elaborated in the introductory section, we aim for short-horizon price scenarios that characterize longer periods sufficiently well to enable substantiated generalization and thus decision-making, e.g., process design assessment.
Thereby, we target case studies with complex process models, for which long scheduling horizons are computationally prohibitive \citep{Schulze.2024}.

We use the DA and ID price scenarios of the suggested \textit{Nominal} and \textit{Unscaled} cases and compare to scenarios generated by clustering methods.
To this end, we examine the economic generalization error from a generated scenario to the detailed results obtained for one year of real (historical) price data.
We then compare the generalization errors from the different price scenarios.

\subsection{Problem description}\label{sec:problem}
We investigate chlor-alkali electrolysis (CAE), which is a classical DR candidate due to high electricity demand \citep{Roh.2019,Bree.2020}.
To facilitate scheduling flexibility, we include a downstream product storage tank.
Given a fixed daily production volume,
we aim to minimize electricity costs over a scheduling horizon. 
To this end, we consider the three scheduling setups where 
(i) only the DA spot market is considered, 
(ii) DA and ID market are included simultaneously, and 
(iii) a two-stage program considering a DA market in the first stage and  ID scheduling subsequently. 
Setup (ii) represents the idealistic case of an exact ID price forecast on the day ahead.
On the other hand, setup (iii) accounts for the chronological order of participation in the two markets.

The scheduling program is formulated as a linear program (LP).
In all three cases, we solve deterministic scheduling LPs of the generic form:
\begin{subequations}
\label{eq:optprob}
\begin{align}
    &\min_{\bm u_t \in\Omega}\hspace{1ex} 
    \frac{1}{T}
    \sum_{t=1}^{T} 
    \Phi_{t}(\bm x_t, \bm u_t)\\
    \mathrm{s.t.}   \hspace{4ex} 
    &\bm f (\bm x_t,\bm x_{t-1},\bm u_t) = \bm 0\,,\;
    \bm x_{0} = \hat{\bm x} \,,\;
    t \in \{1,...,T\} \,,
    \label{eq:process-model}\\
    & \bm c (\bm x_t, \bm x_{t-1}, \bm u_t) \leq \bm 0\,, \;
    t \in \{1,...,T\} \,, \label{eq:constraints}\\
    & \bm x_{k} = \bm x_{0}\,,\; 
    k= 96, 192,..., T \,,
    \label{eq:termcon}
\end{align}
\end{subequations}
wherein $t$ is the time index of quarter-hourly discretization steps on the scheduling horizon, and $T = 96$ for a single-day horizon and $T = 35\,040$ for a full year.
The scheduling degrees of freedom (DOFs), $\bm u_t \in \Omega$, i.e., the traded energy volume(s) $P_t$, are constrained to an admissible set $\Omega \subseteq \mathbb R^{n_u}$.
The states $\bm x_t \in \mathbb{R}^{n_x}$ describe the quater-hourly production volumes $F_t$, and mass production rate $M_t$, and the energy storage $S_t$ (i.e., power integrated over the \unit[15]{min} between $t-1$ and $t$).
Equation \eqref{eq:process-model}, wherein $\bm f: \mathbb{R}^{n_x} \times \mathbb{R}^{n_u} \rightarrow \mathbb{R}^{n_x}$, represents the linear process model.
The dependence on $\bm x_{t-1}$ is due to the integrating storage system.
Equation \eqref{eq:constraints}, wherein
$\bm c: \mathbb{R}^{n_x} \times \mathbb{R}^{n_u} \rightarrow \mathbb{R}^{n_c}$, represents the linear path constraints and ramping constraints on $P_t$, and the linear constraints on $M_t$, $F_t$ and $S_t$.
The ramping constraints reflect that the process cannot realize arbitrarily large steps in the production rate.
Linear constraints on the power purchases from the DA market reassure compliance with the market limitations, i.e., power can be bought up to a maximum allowance and not sold to the DA market.
Finally, Equation \eqref{eq:termcon} represents a periodic constraint on storage at the end of each day, and Equation \eqref{eq:process-model} gives the fixed initial state $\hat{\bm x} \in \mathbb R^{n_x}$.
The detailed equations and parametric values of the process model and problem constraints are taken from \cite{Germscheid.2022}.

To calculate the running electricity cost $\Phi_{t}$, we follow the notation of electricity prices,
$DA_{\lceil t/4 \rceil}$ and $ID_{t}$,
from the preceding sections.
Correspondingly, the purchased DA and ID energy volumes are denoted by $P^\mathrm{DA}_{t}$ and $P^\mathrm{ID}_{s,t}$.
The available scheduling DOFs as well as the calculation of $\Phi_{t}$ differ across the three cases:
\begin{enumerate}
    \item[(i)] $\Phi_{t}^{(\mathrm i)} =DA_{\lceil t/4 \rceil} \cdot P^\mathrm{DA}_{t}$ and $\bm u_t = P^\mathrm{DA}_{t}$.
    \item[(ii)] $\Phi_{t}^{(\mathrm{ii})} =DA_{\lceil t/4 \rceil} \cdot P^\mathrm{DA}_{t} + ID_{t} \cdot P^\mathrm{ID}_{t}$ and $\bm u_t = \{P^\mathrm{DA}_{t}, P^\mathrm{ID}_{t}\}$.
    \item[(iii)] First stage: $\Phi_{t} = \Phi_{t}^{(\mathrm i)}$ and $\bm u_t = P^\mathrm{DA}_{t}$.
    Second stage: $\Phi_{t} = \Phi_{t}^{(\mathrm{ii})}$ and $\bm u_t = P^\mathrm{ID}_{t}$ for $P^\mathrm{DA}_{t}$ fixed.
\end{enumerate}
We assume that $P^\mathrm{DA}_{t}$ is allowed to vary sub-hourly.
When considering multiple price data clusters ($k>1$), we solve Problem \eqref{eq:optprob} for each cluster $s$, $s \in [1,k]$, using the cluster representative prices $DA_{\lceil t/4 \rceil}^s$ and $ID_{t}^s$. 
Then, we weight the cost terms $\Phi_{t}^{s}$ based on cluster size by using the respective cluster weights $\lambda_s$:
\begin{equation} \label{eq:WDC}
    \mathrm{WDC} = 
    \sum_{s=1}^{k} 
    \frac{24 \lambda_{s}}{T}
    \sum_{t=1}^{T}
    \Phi_{t}^{s}\,,
\end{equation}
This weighted daily cost (WDC) is then used to compare the different setups.

\subsection{Price scenario generation}
\label{sec:scenario-generation}
\vspace{-1ex}
We investigate different price scenarios, namely the proposed \textit{Nominal} and \textit{Unscaled} profiles, and scenarios from $k$-means clustering, $k$-medoids clustering, and hierarchical clustering.
Following \cite{Kotzur.2018} and \cite{Xu.2016}, the DA clustering criteria are either (a) the daily mean of the DA prices or (b) the equally-weighted daily mean and daily standard deviation of the DA prices.
For the hierarchical clustering, we consider both medoids and centroids as cluster representators.
The corresponding generation methodologies are termed ``hierarchical-m'' and ``hierarchical-c'', respectively.
In market setups (ii) and (iii), we cluster regarding (c) the daily mean of the DA prices and the daily standard deviation of the ID-DA market deviation, again equally weighted.
This clustering strategy is based on previous findings on price component correlation and frequency analysis \citep{Rahimiyan.2016,Germscheid.2022}.
Here, we focus on comparing the DR case study results for different price generation methods rather than a detailed analysis of the clustering techniques.
Therefore, a limited number of clustering configurations is considered, omitting optimization over a different number of clustered profiles (or clusters) within the same clustering technique.
Further exploration of clustering configurations for price generation in a DR setup is of interest and left for future work.
In total, we here analyze 63 price profiles in 33 price scenarios, numbered S1 to S33, as detailed in Tables \ref{tab:Opti} to \ref{tab:Optiii}.

\subsection{Implementation}
\vspace{-1ex}
We use the Python implementation of the $k$-means, $k$-medoids and hybrid clustering algorithms available in the \texttt{sklearn} \citep{sklearn} package.
The number of clusters considered results from application of the elbow method implemented in the python package \texttt{kneed} \citep{Satopaa.2011}.
We implement Problem~\eqref{eq:optprob} in \texttt{GAMS} and solve the problem using \texttt{CPLEX} with default solver settings.
Similar to Section \ref{sec:Results}, we use historic DA and ID$_3$ price data of the year 2023 from EPEX Spot.

\subsection{Results}
We discuss the case study results for the market setups (i) to (iii).
Based on these results, we extract general conclusions in Section \ref{sec:discussion}.

\subsubsection{Setup i: Scheduling on DA market}
\vspace{-1ex}
Table~\ref{tab:Opti} (upper part) shows that all scenario generation techniques produce results close to the WDC baseline of \unit[5128]{\euro}.
The maximum (absolute) deviation of $\unit[-2.1]{\%}$ occurs in S8,
meaning that the schedule underestimates the average energy costs of one year of operation.
In general, multi-criterion clustering outperforms single-criterion clustering, which highlights the importance of considering both standard deviation and absolute prices.
The schedules obtained when using price scenarios from $k$-means clustering (S4), hierarchical clustering with centroids (S10 and S11), and the \textit{Unscaled} profile (S3) result in higher costs than for the full year (S1).
We explain this loss by the data smoothing associated with profile averaging. 
The results confirm literature observations that $k$-medoids outperform other clustering techniques \citep{Kotzur.2018,Teichgraeber.2019}.
Our \textit{Nominal} DA profile yields $\unit[-0.6]{\%}$ WDC error, which is comparable to $k$-medoids.

 \begin{figure}[htb]
    \vspace{2ex}
	\centering  
 \includegraphics[width=0.92\linewidth]{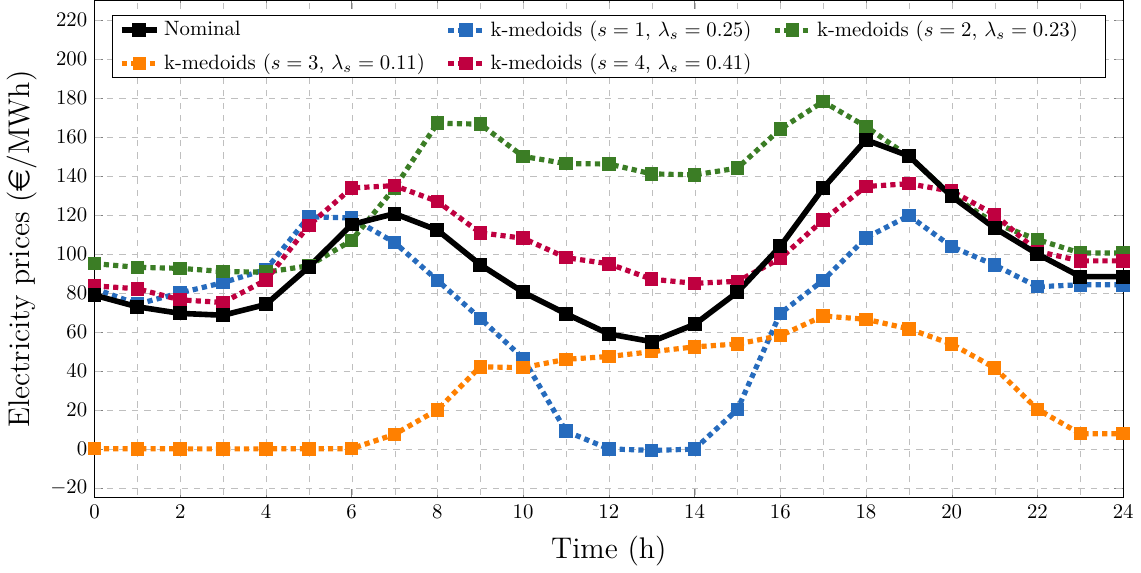}
	\caption{Single-day \textit{Nominal} profile (S2) versus profiles from $k$-medoids clustering (S7).}
 \label{fig:ClusteringPrices}
\end{figure}
For visual comparison, we illustrate the single-day profiles of $k$-medoids multi-criterion clustering (S7) and the \textit{Nominal} profile (S2) in Fig.~\ref{fig:ClusteringPrices}.
The \textit{Nominal} profile lies close to the dominant cluster ($s=4$). 
The corresponding scheduling power profile results are given in the SI.

In some sense, our average-scale-shift strategy is loosely related to clustering techniques.
Hence, to empirically examine similarities and differences, we next compare to single-cluster profiles (lower part of Table \ref{tab:Opti}).
Comparing the centroid-based scenarios (S12, S13, S18, S19) and our (non-scaled and non-shifted) \textit{Unscaled} profile reveals an exact match,
which is not surprising due to the averaging of all data for derivation of the single cluster representator.
Conversely, single-day scheduling based on our \textit{Nominal} profile notably outperforms all other results, highlighting its high potential despite simplicity of the method.
Additionally, we notice that for $k=1$, the $k$-medoids clustering (S14 and S15) performs significantly worse than $k$-means or hierarchical clustering.
Finally, we observe no systematic bias in WDC deviation between single- and multiple-profile generation.
The absence of bias stems from variation in clustering parameters leading to profiles that may or may not capture outliers of the full-year price data.
We highlight that the existence of price outliers, combined with active path constraints and periodical terminal constraints, as outlined in Section \ref{sec:problem}, introduces variability in the results.
In a more complex DR setup or a nonconvex DR optimization problem, this variability becomes even more pronounced, leading to less predictable deviation in the results.

\begin{table}
    \centering
    \caption{Results of Case i. The percentages state the relative deviation of WDC from the benchmark (S1).
    Clustering criteria (a) and (b) as described in Section \ref{sec:scenario-generation}.}
    \label{tab:Opti}
\begin{tabular}{lll ccc}
    \toprule
    &\textbf{Generation} & \textbf{Crit.}& \textbf{k}& \textbf{WDC (\euro)}& \textbf{Dev.~(\%)} \\
    \midrule
    S1 &Full year & $-$ & $-$ & 5128 & $\phantom{-\,}0.0$\\
    S2 &\textit{Nominal} & $-$ & $-$ & 5099 &$-\,0.6$\\
    S3 &\textit{Unscaled} & $-$ & $-$ & 5193&$\phantom{-\,}1.3$\\
    S4 & $k$-means & (a) & 4 & 5186 &$\phantom{-\,}1.1$\\
    S5 & $k$-means & (b) & 4 & 5186 &$\phantom{-\,}1.1$\\
    S6 &$k$-medoids & (a) & 5 & 5180 &$\phantom{-\,}1.0$\\
    S7 & $k$-medoids & (b) & 4 & 5129 & $\phantom{-\,}2\cdot 10^{-2}$\\
    S8 & Hierarchical-m & (a) & 4 & 5021 & $-\,2.1$\\
    S9 & Hierarchical-m & (b) & 4 & 5067 & $-\,1.2$\\
    S10 &Hierarchical-c & (a) & 4 & 5185 & $\phantom{-\,}1.1$\\
    S11 & Hierarchical-c & (b) & 4 & 5188 & $\phantom{-\,}1.2$\\
    \midrule
    S12 &$k$-means & (a) & 1 & 5193 &$\phantom{-\,}1.3$\\
    S13 & $k$-means & (b) & 1 & 5193 &$\phantom{-\,}1.3$\\
    S14 &$k$-medoids & (a) & 1 & 5299 &$\phantom{-\,}3.3$\\
    S15 & $k$-medoids & (b) & 1 & 5423 & $\phantom{-\,}5.7$\\
    S16 &Hierarchical-m & (a) & 1 & 5248 & $\phantom{-\,}2.3$\\
    S17 & Hierarchical-m & (b) & 1 & 5110 & $-\,0.4$\\
    S18 &Hierarchical-c & (a) & 1 & 5193 & $\phantom{-\,}1.3$\\
    S19 & Hierarchical-c & (b) & 1 & 5193 & $\phantom{-\,}1.3$\\
    \bottomrule    
\end{tabular}
\end{table}

\subsubsection{Setup ii: Simultaneous scheduling on DA \& ID market}
We present the scheduling results in Table~\ref{tab:Optii}.
As expected, simultaneous participation in the DA and ID markets reduces costs compared to the DA-only setting. 
The cost deviations from the baseline are higher (up to \unit[9.4]{\%}) than in Case (i),
with the largest errors for centroid-based clustering (S23, S26) and our \textit{Unscaled} profile (S22).
We attribute the higher deviations to a high error sensitivity in the interplay of two price profiles.
Again, we observe substantial deviations for centroid-based clustering (S23, S26) and our \textit{Unscaled} profile (S22).
In contrast, using our \textit{Nominal} DA and ID profiles (S21) only yields slightly larger errors than the medoid-based ($k=6,4$) multi-scenarios (S24 and S25), and give an absolute error of \unit[0.6]{\%} over scheduling based on the full year profiles.

\begin{table}
    \centering
    \caption{Results of Case ii. The percentages state the relative deviation of WDC from the benchmark (S20). S23 to S26 use clustering criteria (c) as described in Section \ref{sec:scenario-generation}.}
\begin{tabular}{lll cc}
    \toprule
    &\textbf{Generation}& \textbf{k}& \textbf{WDC (\euro)}& \textbf{Dev.~(\%)} \\
    \midrule
    S20 & Full year & $-$ & 4393 & $\phantom{-\,}0.0$\\
    S21 & \textit{Nominal} & $-$ & 4318 & $-1.7$\\
    S22 & \textit{Unscaled} & $-$ & 4798 &$\phantom{-\,}9.2$\\
    S23 & $k$-means & 5 & 4805 & $\phantom{-\,}9.4$\\
    S24 & $k$-medoids & 6 & 4367 &$-0.6$\\
    S25 & Hierarchical-m & 4 & 4369 & $\phantom{-\,}0.6$\\
    S26 & Hierarchical-c & 4 & 4808 &  $\phantom{-\,}9.4$\\
    \bottomrule    
\end{tabular}\label{tab:Optii}
\end{table}

\subsubsection{Setup iii: Two-stage scheduling on DA \& ID market}
Table \ref{tab:Optiii} collects the scheduling results of the two-stage problem. 
Here, the values of WDC and relative deviation fall between Cases i and ii.
The best results are obtained when using our \textit{Nominal} DA and ID profiles (S28), closely followed by the medoid-based hierarchical method (S33) and $k$-medoids clustering (S31).
As before, centeroid-based clustering (S30 and S33) and our \textit{Unscaled} profiles (S29) provide the least accurate prediction of energy costs.
We found similar results when considering a one-week horizon, i.e., $T = 672$ (see SI).

\begin{table}[htb]
    \centering
        \caption{Results of Case iii. The percentages state the relative deviation of WDC from the benchmark (S20). S30 to S33 apply clustering criteria (c) as described in Section \ref{sec:scenario-generation}}
\begin{tabular}{lll cc}
    \toprule
    &\textbf{Generation}& \textbf{k}& \textbf{WDC (\euro)}& \textbf{Dev.~(\%)} \\
    \midrule
    S27 & Full year & $-$ & 5029 & $\phantom{-\,}0.0$\\
    S28 & \textit{Nominal} & $-$ & 4996 & $-0.6$\\
    S29 & \textit{Unscaled} & $-$ & 5152 & $\phantom{-\,}2.4$\\
    S30 & $k$-means & 5 & 5136 & $\phantom{-\,}2.1$\\
    S31 & $k$-medoids & 6 & 4968 & $-1.2$\\
    S32 &Hierarchical-m & 4 & 5064 & $\phantom{-\,}0.7$\\
    S33 & Hierarchical-c & 4 & 5135 & $\phantom{-\,}2.1$\\
    \bottomrule    
\end{tabular}\label{tab:Optiii}
\end{table}

\subsection{Discussion}
\label{sec:discussion}
Throughout all scheduling setups (i) to (iii), we have calculated cost errors for the consideration of the proposed price generation and clustering techniques to the use of full-year data. We have shown that
employing a single \textit{Nominal} profile scenario facilitates accurate prediction and generalization of economic savings within a DSM scheduling strategy.
Notably, in Case i, using our \textit{Nominal} scenario outperformed state-of-the-art $k=1$ clustering methods (except hierarchical-m clustering with similar results).
More generally, the predictions with our price scenario generation strategy consistently belongs to the two most accurate result throughout all cases, having generalization errors below $\pm\,\unit[2]{\%}$. 
This finding supports the use of (variance) scaling within profile construction.

Considering that DSM scheduling savings commonly lie between \unit[2]{\%} and \unit[10]{\%} \citep{Otashu.2018,Caspari.2020,Mucci.2023},
we regard scenario generalization errors above \unit[2]{\%} as economically prohibitive.
Hence, our scenario generation method produces acceptable generalization errors in the present study.
This indicates that a single representative profile can be sufficiently descriptive for early-stage assessment of processes and operation strategies  even outperforming multi-scenario clustering methods.

Among all clustering techniques, the medoid-based methods (namely $k$-medoids and hierarchical clustering with a medoid-derived center profile) produce the most accurate scheduling results regarding generalization from a single day to the entire year.
Further, considering multiple (here two) clustering criteria over a single criterion yields superior results.
Both observations agree with the literature \citep{Teichgraeber.2019,Kotzur.2018}.

When considering a multi-stage problem (iii), the \textit{Nominal} profile proves to be more effective than clustering techniques. 
In that case, multi-criteria clustering struggles with complex criteria selection and inconsistent representation of individual market behaviors.
On the other hand, clustering for each market separately would disregard crucial dependencies between the prices as discussed by \cite{Schafer.2020}, possibly preventing conclusions about the proposed DR method.
Our method overcomes these challenges by retaining key statistical values, namely the average of the DA profile, the standard deviation of both DA and ID profiles, and the dominant fluctuation patterns of both profiles.
Moreover, considering the DA-ID deviation allows for more substantial conclusions and facilitates a zero-integral constraint on the constructed price tuple.

\section{Conclusions}
We present a method to construct electricity price scenarios based on an averaging and scaling procedure using historical price data.
Our approach is capable of generating both DA and ID scenarios.
To reassure consistency, we adjust the cumulative deviation of a DA-ID pair to zero-mean deviation.
Our method is constructed to be simple and intuitive, to have few adjustable parameters, and thus be easily reproducible.
At the same time, the method can provide sufficiently realistic price profiles, by utilizing recent price data.
In particular, we showed that the \textit{Nominal} profile generation strategy facilitates the generalization from a single-day scheduling horizon to longer time periods, e.g., a full year.
Overall, we aim to ease the selection of a price profile in future case studies on DSM.

We have applied our method to historical DA and ID data from EPEX spot for the year 2023.
Therein, we generated multiple price scenarios with varying magnitude of price fluctuation.
These profiles are ready-to-use and openly accessible\footnote{https://git.rwth-aachen.de/avt-svt/public/representative-electricity-price-profiles}.
The suggested DA and ID profiles closely represent historical time-series of DA and ID profiles realized on the same day.
In a DR case study considering DA and ID markets, we compared the suggested price scenarios to single- and multi-clustering techniques.
Therein, our suggested \textit{Nominal} price scenarios provided consistently accurate prediction, comparable to medoid-based multi-scenario hierarchical clustering, while outperforming all other clustering methods.
Further, we have recently presented the first application of the price scenarios in a more complex integrated scheduling and control case study \cite{Schulze.2024}.

We acknowledge that some advanced scenario generation methods additionally consider higher-order statistical moments in their generation procedure and might lead to even more accurate results.
However, these approaches are inherently more involved, result in a greater number of adjustable parameters, and can trigger considerably high computational times.
In general, both historical data-based and scenario forecasting methods have been inaccurate at predicting the long-term evolution of prices.
Clearly, this long-term evolution is subject to many factors, including trading policy and political changes.
The forecast prices additionally refer to a specific forecasting time period and do not serve for generalizable results, e.g., they are time-specific and do not provide an average basis for evaluating the broader applicability when benchmarking a new DR concept.

Hence, we focus on approaches to reflect the current character of the electricity price on spot markets.
Since we are interested in an intuitive and deterministic method rather than covering all possible price profile shapes, we accept to trade such a flexibility for the desired simplicity of the proposed averaging-based method.

{\footnotesize \textbf{Acknowledgements.}
The authors gratefully acknowledge the financial support of the Kopernikus project SynErgie by the Federal Ministry of Education and Research (BMBF), and the project supervision by the project management organization Projekttr\"ager J\"ulich (PtJ). 
The authors gratefully acknowledge the financial support of and the Deutsche Forschungsgemeinschaft (DFG, German
Research Foundation)—333849990/GRK2379 (IRTG Hierarchical and Hybrid Approaches in Modern Inverse Problems).
Moreover, we thank Simone Mucci for fruitful discussions as well as Sonja Germscheid and Eike Cramer for constructive feedback on the manuscript.}

\clearpage
\bibliography{Bibs}

\end{document}